\documentclass[11pt,a4paper]{article}
\usepackage{jheppub}                % See geometry.pdf to learn the layout options. There are lots.

\usepackage{epsfig,multicol,bbm}
\usepackage{amsmath,amssymb,bm}
\usepackage{graphicx}

\title{Hybrid Monte Carlo on Lefschetz Thimbles  \\ -- A study of the residual sign problem -- }
\author[]{H.~Fujii,}
\author[]{D.~Honda,}
\author[]{M.~Kato,}
\author[]{Y.~Kikukawa,}
\author[]{S.~Komatsu}
\author[b]{and T.~Sano}
\affiliation[]{Institute of Physics, the University of Tokyo, Tokyo 153-8902, Japan }
\affiliation[b]{Theoretical Research Division, RIKEN Nishina Center, Wako 2-1, Saitama 351-0198, Japan}

\emailAdd{hfujii@phys.c.u-tokyo.ac.jp}
\emailAdd{dhonda@hep1.c.u-tokyo.ac.jp}
\emailAdd{kato@hep1.c.u-tokyo.ac.jp}
\emailAdd{kikukawa@hep1.c.u-tokyo.ac.jp}
\emailAdd{skomatsu@hep1.c.u-tokyo.ac.jp}
\emailAdd{tsano@riken.jp}

\subheader{\rm \small \hfill UT-KOMABA/13-11}

\abstract{
We consider a hybrid Monte Carlo algorithm which is applicable to lattice theories defined 
on Lefschetz thimbles. 
In the algorithm, any point (field configuration) on a thimble is parametrized uniquely by the flow-direction 
and the flow-time defined at a certain asymptotic region close to the critical point,  
and it is generated by solving the gradient flow equation downward.
The associated complete set of tangent vectors is also generated in the same manner. 
Molecular dynamics is then formulated as a constrained dynamical system, where the equations of motion 
with Lagrange multipliers are solved by the second-order constraint-preserving symmetric integrator. 
The algorithm is tested in the $\lambda \phi^4$ model at finite density, by choosing the thimbles associated 
with the classical vacua for subcritical and supercritical values of chemical potential. 
For the lattice size $L=4$,  we find that the residual sign factors average to not less  than 0.99 and 
are safely included by reweighting and that the results of the number density are consistent 
with those obtained by the complex Langevin simulations.
}
\keywords{Lattice Field Theory, Monte Carlo simulation, Lefschetz thimble}
\begin{document}
\maketitle
%\section{}
%\subsection{}

\section{Introduction}
\label{sec:introduction}
Formulated in physically 
well-reasoned and well-defined manners, 
several field  theories have complex actions in Euclidean lattice. %space-time.
These include QCD at finite density,  chiral gauge theories,  chiral Yukawa theories, etc.
To these theories, the state-of-art Monte Carlo methods %for nonperturbative study 
do not apply straightforwardly.
If there exists a stochastic method which 
is based on  a sound theoretical basis and is applicable to such theories with complex actions,  
it would allow us  to do thorough non-perturbative studies of these theories.\footnote{In particular, 
the study of lattice QCD at finite temperature and density, 
to figure out the phase structure of QCD,  is the subject of great interest
and there are a lot of research activities. 
The authors refer the reader to 
the talks at the International Workshop on the Sign Problem in QCD and Beyond 
(Regensburg, 2012)[{\tt http://www.physik.uni-regensburg.de/sign2012/talks.shtml}], 
and \cite{deForcrand:2010ys, Gupta:2011ma, Levkova:2012jd, Ejiri:2013lia, Aarts:2013bla,Gattringer-lattice2013-plenary} for recent reviews and references.
} 

One possible % and, in a sense, natural 
approach to this problem is to consider the field variables, which are assumed to be real in the original formulation, to be complex and to extend the cycle of path-integration to a complex region in order to 
achieve better convergence. First example is to use complexified Langevin 
equation\cite{Parisi:1984cs,Klauder:1983zm,Klauder:1983sp}. Second example is to select 
Lefschetz thimbles as the cycle of path-integration,  where the imaginary part of the complex action 
stays constant\cite{Witten:2010cx,Cristoforetti:2012su}.

The first method,  to use complexified 
Langevin equation\cite{Parisi:1984cs,Klauder:1983zm,Klauder:1983sp},
%This method 
is simple in its implementations, but it is not fully understood theoretically in its convergence properties\cite{Ambjorn:1985iw,
Ambjorn:1986fz, 
Berges:2006xc, 
Berges:2007nr,
Aarts:2008rr,
Aarts:2008wh,
Aarts:2009hn,
Aarts:2009dg,
Aarts:2009uq, 
Aarts:2010aq, 
Aarts:2011zn,
Aarts:2011ax, 
%Fromm:2012eb,
Seiler:2012wz, 
Pawlowski:2013pje, 
Pawlowski:2013gag, 
Aarts:2013uxa,
Aarts:2013uza,
Sexty:2013ica,
Aarts:2013fpa, 
Giudice:2013eva}.\footnote{See  \cite{Aarts:2013bla,Aarts:2013uxa} for reviews on this approach.}
The recent numerical results by Aarts  about the complex 
$\lambda \phi^4$ model%at finite density
 \cite{Aarts:2008wh}\footnote{This model has also been simulated 
successfully by the dual variable method in \cite{Gattringer:2012df,Mercado:2012ue,Gattringer:2012jt}.}
and 
by Sexty on full QCD\cite{Sexty:2013ica}  % at nonzero density
are remarkable and encouraging, though.
%is , showing the better convergence by the use of gauge cooling. 

The second method, 
to select Lefschetz thimbles as the cycle of path-integration\cite{Witten:2010cx,Cristoforetti:2012su,Cristoforetti:2012uv,Cristoforetti:2013wha}, 
%where the imaginary part of complex actions stay constant, 
%This method 
seems generic, but it is not easy in general to know 
the set of thimbles which is equivalent to the original cycle.
Moreover,  
the path-integration measure in the complexified field space gives rise to an extra complex phase, 
and  to compute the residual phase factor it is required to know the tangent spaces of the thimbles. 
Recently, AuroraScience collaboration\cite{Cristoforetti:2012su}  has considered 
to define a lattice model by the single thimble associated with the Gaussian critical point (or the classical vacuum) and 
proposed a Langevin simulation algorithm for such models.\footnote{It has also been shown by the authors of \cite{Cristoforetti:2012su} that in such models
the symmetry property is preserved and the perturbation theory reproduces the same result as the original one.} The collaboration has then reported  a numerical result about the $\lambda \phi^4$ model \cite{Cristoforetti:2013wha},  
which is consistent with the results obtained by the complex Langevin equation\cite{Aarts:2008wh} 
and the dual variable method\cite{Gattringer:2012df,Mercado:2012ue,Gattringer:2012jt}.\footnote{Strictly speaking, the field configurations obtained in this simulation do not belong to the 
Lefschetz thimble associated with the Gaussian critical point (or the classical vacuum).
% in the case with the external source $h$). 
Rather, they are obtained by projecting 
on to the tangent space at the critical point, 
although the tangent space is not necessarily in the same homology class as the thimble. 
Accordingly, the imaginary part of the action does not stay constant, and its exponent is included by reweighting.
It has been claimed that this approximation is good enough to reproduce 
the silver blaze behavior in the $\lambda \phi^4$ model. 
}
In these works, however, the residual phase factor is ignored, and
the residual sign problem in the $\lambda \phi^4$ model remains to be studied systematically. 
Quite recently, 
Mukherjee, Cristoforetti and Scorzato have studied the residual sign problem in the U(1) one-plaquette model through 
a new Metropolis sampling method\cite{Mukherjee:2013aga}.\footnote{This Metropolis 
sampling method is
based on a mapping between a thimble and its asymptotic ``Gaussian" region close to the critical point. }
%In order to clarify the usefulness of this approach, however, it is desirable to study  systematically the residual sign problem. 

%To traverse Lefschetz thimbles, it is required to know about the tangent space of the thimbles.
%Also the residual sign factor is determined by Jacobian associated the coordinate transformation to the tangent space.

The purpose of this article is to 
%In this article we 
introduce a hybrid Monte Carlo algorithm 
%for 
which is applicable to 
the lattice models defined on Lefschetz thimbles. 
In this algorithm, 
any point (field configuration) on a given thimble is 
parametrized uniquely by the flow-direction and the flow-time defined in a certain asymptotic 
region close to the critical point, % using the fourth-order Runge-Kutta method. 
and it is generated 
by solving  the gradient flow equation downward.
The complete set of tangent vectors at the point (associated with the field configuration) is also generated, although it is numerically very demanding, but because it 
%and the residual sign factor is computed using it.
%which span the tangent space  (the space of the infinitesimal variation of the field configuration)
%of the vicinity of the filed configuration
is required for the computation of the residual phase factor.
Molecular dynamics %to traverse a thimble 
is then formulated 
as a constrained dynamical system, where
the equations of motion with Lagrange multipliers are solved 
by the second-order constraint-preserving symmetric integrator\cite{Leimkuhler-Reich}.
%Using the set of tangent vectors,  the residual sign factor is computed. 
%
%This algorithm, we hope, may be 
We hope this algorithm can be used for a systematic study of the residual sign problem
and other aspects of this second method.
%of the usefulness of the second method.

We test the algorithm in the complex $\lambda \phi^4$ model at finite density by observing 
%averages of 
the number density 
%and the residual sign factor  
for various values of chemical potential  $\mu$. 
We examine both thimbles 
associated with 
the classical vacua for subcritical and supercritical values of $\mu$.
%
%associated with symmetric classical vacuum  for subcritical $\mu$ and
%associated with broken classical vacuum  for supercritical $\mu$.  
For 
%the couplings $\kappa=1$ and $\lambda=1.0$, 
%the chemical potential $\mu \in [0.0, 1.5]$($\mu_c \simeq 0.962$), and 
the lattice size $L=4$, 
%For the hypercubic lattice %$\mathbb{L}^4$ 
%with the linear extent $L=4$ and the chemical potential $\mu$ within the range $[0,1.5]$,  
we find that the residual sign factors average to not less  than 0.99 and are safely 
included by reweighting, 
and that 
the results of the number density agree with those obtained by the complex Langevin simulations 
within statistical errors, except for a few values of $\mu$, 
and overall, they are consistent with each other.
% in the range $\mu \in [0, 1.5]$

This paper is organized as follows.
In section~\ref{sec:complexification}, we review the basics of the complexification of lattice models on Lefschetz thimbles.
Section~\ref{sec:HMC-on-Lefschetz-thimble} is devoted to the description of the hybrid Monte Carlo algorithm which is applicable to lattice models defined on Lefschetz thimbles.
In section~\ref{sec:test}, the algorithm is applied 
to the $\lambda \phi^4$ model with chemical potential. % $\mu$. 
In the final section~\ref{sec:summary-discussion},  we conclude with a few discussions.

\section{Complexified models on Lefschetz thimbles}
\label{sec:complexification}
%In this section, we first 
First we %discuss
review the basics of  
the complexification of lattice models on Lefschetz thimbles\cite{Witten:2010cx,Cristoforetti:2012su}.
Let us consider a lattice theory with  $n$ real degrees of freedom and denote  the real field variables  
%as $x_i $$(i=1,\cdots,n)$ and  
as $x=(x_1,\cdots, x_n)$. % collectively. 
It is assumed that $x$ takes the value in a subset $\mathcal{C}_{\mathbb{R}}$ of $\mathbb{R}^n$
and the action of the model $S[x]$ has a non-zero imaginary part. 
%\begin{eqnarray}
%S[x] = {\rm Re} \, S[x] + i {\rm Im} \, S[x] , \qquad  x \in C_{\mathbb{R}}  \, (\subseteq \mathbb{R}^n ).
%\end{eqnarray}
The partition function of the model is defined by the path-integration 
over $\mathcal{C}_{\mathbb{R}}  \, (\subseteq \mathbb{R}^n )$, 
\begin{equation}
Z = \int_{{\mathcal C}_\mathbb{R}}{\cal D}[x]  \exp\{ - S[x]  \}, 
\end{equation}
where the measure is given by ${\cal D}[x] =d^n x$.
%\footnote{A non-trivial  measure term, if any, may be included into ${\cal D}[x]$ or into the action $S[x]$.} 

In complexification, 
%
%Next we consider a complexification of the model:
%The model is complexified:
the field variables  are extended to complex variables $z \in \mathbb{C}^n$, 
and the action is extended to a holomorphic function of $z$,  $S[z]$. %$S[x] \rightarrow S[z]$. 
As for the cycle of the path-integration,  
Morse theory %(Picard-Lefschetz theory) 
tells us how to select the set of Lefschetz thimbles which is homologically equivalent to 
${\mathcal C}_\mathbb{R}$. %as  a cycle of the path-integration.
Morse function in our case is defined by $h \equiv - {\rm Re} \,  S [ z]$ and the associate 
gradient (downward) 
flow equation is given by\footnote{Along the flow, $h$ is monotonically decreasing, 
\begin{equation*}
\frac{d}{dt} h = - \frac{1}{2} \left\{ \frac{ \partial S [ z] }{\partial z}  \cdot \frac{d}{dt} z(t) 
                                                       +\frac{ \partial \bar S [ \bar z] }{\partial \bar z}  \cdot \frac{d}{dt} \bar z(t)
\right\}
                      = - \left\vert \frac{ \partial S [ z] }{\partial z} \right\vert^2   \le 0 , 
\end{equation*}
while the imaginary part of the action stays constant, 
\begin{equation*}
\frac{d}{dt}  {\rm Im}\, S[z] = \frac{1}{2i}\left\{ \frac{ \partial S [ z] }{\partial z}  \cdot \frac{d}{dt} z(t) 
                                                       -\frac{ \partial \bar S [ \bar z] }{\partial \bar z}  \cdot  \frac{d}{dt} \bar z(t)
\right\}
=0 . 
\end{equation*}
}
\begin{eqnarray}
\label{eq:downward-flow-equation}
\frac{d}{dt} z_i(t) =  \frac{ \partial \bar S [ \bar z] }{\partial \bar z_i} 
%, \qquad
%\frac{d}{dt} \bar z(t) =  \frac{ \partial S [ z] }{\partial z}
\qquad  
( t \in \mathbb{R} ).
\end{eqnarray}
%\begin{eqnarray}
%\label{eq:downward-flow-equation}
%\dot{z}_i (t) = \bar \partial_i \bar S [ \bar z] 
%%\frac{d}{dt} z_i(t) =  \frac{ \partial \bar S [ \bar z] }{\partial \bar z_i} 
%%, \qquad
%%\frac{d}{dt} \bar z(t) =  \frac{ \partial S [ z] }{\partial z}
%\qquad  
%( t \in \mathbb{R} ). 
%\end{eqnarray}
The set of critical points %of the flow equation 
$\Sigma$ consists of 
the points $\{ z_\sigma \}$ which satisfy 
$\left.  \partial S [ z ] / \partial \bar z_i \right\vert_{z=z_\sigma}=0$.
Associated with a critical point $z_\sigma$,  
%\begin{eqnarray}
%\left.  \frac{ \partial S [ z ] }{\partial z} \right\vert_{z=z_\sigma}=0, 
%\end{eqnarray}
a Lefschetz thimble $\mathcal J_\sigma$ is defined by 
the union of all downward flows which trace back to $z_\sigma$ at $t=-\infty$.  The thimble is a $n$-dimensional real submanifold in $\mathbb{C}^n$. One can introduce   
another $n$-dimensional real submanifold 
${\cal K}_\sigma$ 
of $\mathbb{C}^n$ by the union of all downward flows which converge to $z_\sigma$ at $t=+\infty$ so that 
its intersection number is unity with  $\mathcal J_\sigma$ and vanishing otherwise,  
$\left\langle {\cal J}_\sigma , {\cal K}_\tau \right\rangle = \delta_{\sigma \tau}  $. 
Then, according to Morse theory,  it follows that 
\begin{equation}
{\mathcal C}_\mathbb{R} 
%= \mathbb{R}^n
= \sum_{\sigma \in \Sigma}  n_\sigma {\cal J}_\sigma , \qquad 
n_\sigma =  \left\langle {\mathcal C}_\mathbb{R} , {\cal K}_\sigma  \right\rangle . 
\end{equation}
And the partition function of the model is given by the formula,
%\begin{equation}
%\label{eq:Z-with-Lefschetz-thimbles}
%Z = 
%\sum_{\sigma \in \Sigma}  n_\sigma  \exp\{ -S[z_\sigma]  \}
%\int_{{\cal J}_\sigma}{\cal D}[z]  \exp\{ - {\rm Re} \big( S[z] - S[z_\sigma] \big)  \}.
%\end{equation}
\begin{eqnarray}
\label{eq:Z-with-Lefschetz-thimbles}
Z &=& 
\sum_{\sigma \in \Sigma}  n_\sigma  \exp\{ -S[z_\sigma]  \} \, Z_\sigma ,  \\
\label{eq:Z-with-Lefschetz-thimbles2}
Z_\sigma &=& \int_{{\cal J}_\sigma}{\cal D}[z]  \exp\{ - {\rm Re} \big( S[z] - S[z_\sigma] \big)  \}.
\end{eqnarray}
In this result,  
for the critical points $\{ z_\sigma \}$ satisfying 
$-  {\rm Re} S[z_\sigma] >  \text{max}\left\{ - {\rm Re} S[x] \right\} (x \in {\mathcal C}_{\mathbb{R}})$, 
it holds that 
$\langle {\mathcal C}_\mathbb{R}, {\cal K}_\sigma \rangle  = 0$  
and the associated thimbles do not contribute to the path-integration. 
On the other hand, 
for the critical points $\{ z_\sigma \}$ in the original cycle ${\mathcal C}_\mathbb{R}$ 
(i.e. classical solutions in the original theory), 
it holds that 
$\langle {\mathcal C}_\mathbb{R}, {\cal K}_\sigma \rangle  = 1$ 
and 
the associated thimbles contribute 
with the relative weights proportional to $\exp ( - S[z_\sigma] )$. 
%$\exp ( - {\rm Re} \, S[z_\sigma] )$. 
In particular,  for the classical vacuum in the original theory
$z_{\rm vac} \in {\mathcal C}_{\mathbb{R}}$, 
it holds that
 $- {\rm Re} \, S[z_{\rm vac}]= \text{max}\left\{ - {\rm Re} S[x] \right\} (x \in {\mathcal C}_{\mathbb{R}})$ and therefore the associated thimble $\mathcal J_{\rm vac}$ contributes most among all the thimbles.
% \footnote{The classical vacuum in the original theory $z_{\rm vac}$ is not necessarily the Gaussian critical point, $z=0$. }
And, in the above formula eq.~(\ref{eq:Z-with-Lefschetz-thimbles2}), 
the measure on the thimbles 
$\left. {\cal D}[z] = d^n z \right\vert_{ {\cal J}_\sigma}$
should be specified based on the knowledge of 
the geometry of $\{ {\cal J}_\sigma \}$, in particular,  their tangent spaces.

As to the expectation value of an observable $O[z]$,  it is defined by the formula,
\begin{equation}
\label{eq:observables-on-thimbles}
\langle O[z] \rangle = \frac{1}{Z} 
\sum_{\sigma \in \Sigma}  n_\sigma  \exp\{ -S[z_\sigma]  \} \, 
Z_\sigma \, \langle O[z] \rangle_{{\cal J}_\sigma}, 
\end{equation}
where
\begin{eqnarray}
%Z_\sigma &=& \int_{{\cal J}_\sigma}{\cal D}[z]  \exp\{ - {\rm Re} \big( S[z] - S[z_\sigma] \big)  \} , \\
\langle O[z] \rangle_{{\cal J}_\sigma} &=& \frac{1}{Z_\sigma} 
\int_{{\cal J}_\sigma}{\cal D}[z]  \exp\{ - {\rm Re} \big( S[z] - S[z_\sigma] \big)  \}  \, O[z] . 
\end{eqnarray}
As a possible and practical approximation to the formula eq.~(\ref{eq:observables-on-thimbles}), 
one may take the single contribution of the thimble associated with 
the classical vacuum, $\mathcal{J}_{\rm vac}$, as considered by 
AuroraScience collaboration\cite{Cristoforetti:2012su}.\footnote{
While $\langle O[z] \rangle_{{\cal J}_\sigma}$ may be evaluated through 
Monte Carlo simulations as discussed in \cite{Cristoforetti:2012su,Cristoforetti:2013wha} and 
will be discussed in the following sections, 
it is not straightforward to 
compute $\{ Z_\sigma \}$$(\sigma \in \Sigma)$ in general. 
At one-loop, i.e. in the saddle point approximation,  
$Z_\sigma = 1/ \sqrt{ \det {K }}$ where ${K }$ 
is defined in eq.~(\ref{eq:downward-flow-equations-linearized}) below. }
In this approximation, the above formula is simplified as follows:
\begin{equation}
\label{eq:observables-on-vacuum-thimble}
\left\langle O[z] \right\rangle = \langle O[z] \rangle_{\mathcal{J}_{\rm vac}} . 
\end{equation}

We then summarize a few geometric properties of Lefschetz thimbles.
%A summary of some of the geometric properties of Lefschetz thimbles is in order.
First we recall that for a given critical point $z_\sigma \in \Sigma$, 
the associated thimble ${\cal J}_\sigma$ is 
the union of all downward flows which trace back to $z_\sigma$ at $t=-\infty$.
In the vicinity of the cirtical point $z_\sigma$, 
the flow equation eq.~(\ref{eq:downward-flow-equation}) 
%$d z(t) / dt = \bar \partial_i \bar S [ \bar z]$
can be linearized as\footnote{In the following, we will use the abbreviation
%$d f(t) / dt = \dot{f}(t)$ and 
$\partial / \partial z_i = \partial_i$, $\bar \partial / \partial \bar z_i =\bar \partial_i$. }
\begin{equation}
\label{eq:downward-flow-equations-linearized}
%\Delta  \dot{z}_i (t) =  \bar K_{ij} \,  \Delta \bar z_j(t),  
\frac{d}{dt} \big(z_i(t) -  {z_\sigma}_i \big) =  \bar K_{ij} \, \big(\bar z_j(t) - \bar z_{\sigma j}\big), 
\qquad K_{ij} \equiv \left. \partial_i \partial_j S [z]  \right\vert_{z=z_\sigma} . 
\end{equation}
%where $K_{ij} \equiv \left. \partial_i \partial_j S [z]  \right\vert_{z=z_\sigma}$. 
The complex symmetric matrix $K_{ij}$,  according to the Takagi factorization theorem\cite{Horn-Johnson}, can be cast into a  positive diagonal matrix as
$v^\alpha_i K_{ij} v^\beta_j = \kappa^\alpha  \delta^{\alpha \beta}$, 
where $\kappa^\alpha \ge 0$ $(\alpha=1,\cdots, n)$ and $v_i^\alpha$$(\alpha=1,\cdots, n)$ are orthonormal complex vectors. And the solution to the  linearized flow equation is obtained as 
\begin{equation}
\label{eq:downward-flow-equations-linearized-solution}
z_i(t) -  {z_\sigma}_i  = v_i^\alpha \,\exp\big( \kappa^\alpha (t-t_0) \big) \, \xi_0^\alpha, \qquad 
\xi_0^\alpha \in \mathbb{R} \, \,(\alpha=1,\cdots, n).
\end{equation}
Indeed, the set of the orthonormal vectors 
$\{ v^\alpha \}$$(\alpha=1,\cdots, n)$
spans the tangent space of the Lefschetz thimble ${\cal J}_\sigma$ at the critical point $z_\sigma$,
$T_{z_\sigma}$:
close to the critical point, the thimble is parametrized 
by $n$ real parameters $\xi^\alpha \in \mathbb{R} \, (\alpha=1,\cdots, n)$ as 
$ z_i - {z_\sigma}_i \simeq v_i^\alpha \, \xi^\alpha $, 
%\begin{equation}
%z_i \simeq \xi^\alpha v_i^\alpha, \qquad \xi^\alpha \in \mathbb{R} (\alpha=1,\cdots, n), 
%\end{equation}
and the action reads 
$S[z] -S[z_\sigma] \simeq (z_i - {z_\sigma}_i) K_{ij} (z_j - {z_\sigma}_j) /2
=  \kappa^\alpha \xi^\alpha \xi^\alpha /2\, \, \in \mathbb{R}$.
%\begin{equation}
%S[z] = \frac{1}{2} z_i K_{ij} z_j = \frac{1}{2}  \vert \kappa^\alpha \vert \xi^\alpha \xi^\alpha \, \,    \in \mathbb{R} .
%\end{equation}

At a generic  point $z$ on the thimble ${\cal J}_\sigma$, 
one can also define a tangent space $T_z$ and
a basis of  tangent vectors $\{ V_z^\alpha \} (\alpha =1,\cdots, n)$. 
Because any two tangent vectors $V_z$ and $V_z^\prime$ should 
commute with each other, 
$\{ V_z \partial + \bar V_z \bar \partial \}  V_z^\prime
- \{ V_z^\prime \partial + \bar V_z^\prime \bar \partial \} V_z =0$, and 
the direction vector of the gradient flow, $g \equiv \bar \partial \bar S[\bar z]$, itself 
should be a tangent vector, 
it follows that
$\{ V_z^\alpha \}$ satisfy the following flow 
equations,\footnote{
The commutation relation of two vectors $V_z^\alpha$ and $V_z^\beta$, if 
one of the vectors is set to  
the direction vector of the Lefschetz flow $g \equiv \bar \partial \bar S[\bar z]$, 
reads $\{ g \partial + \bar g \bar \partial\} V_z^\alpha - \{ V_z^\alpha \partial + \bar V_z^\alpha \bar \partial \} g =0. $ This immediately implies that 
\begin{equation*}
\frac{d}{dt} {V}_{z i}^\alpha(t) 
%= \{ w \partial + \bar w \bar \partial\} V^\alpha \\
=\{ V_z^\alpha \partial + \bar V_z^\alpha \bar \partial \} g_i \\
=  \bar \partial_i  \bar \partial_j \bar S[ \bar z] \, \bar V^\alpha_{z j}(t) . 
\end{equation*}
}
\footnote{
In the vicinity of the critical point $z_\sigma$, the flow equation for the 
tangent vectors eq.~(\ref{eq:downward-flow-equation-tangent-vectors}) is linearized as
$d {V}_i^\alpha(t)/ dt = \bar K_{ij}  \bar V^\alpha_j(t)$. 
And the solution to the equation is obtained as 
\begin{eqnarray*}
V_i^\alpha(t) 
%&\simeq&  v^\beta \exp(\vert \kappa^\beta \vert (t-t_0) ) \, Q^{\beta \alpha}(t_0)  \\
&=&  v_i^\beta \, \exp\big(\kappa^\beta \, (t-t_0) \big) \, C_0^{\beta \alpha}, \qquad
C_0^{\beta \alpha} \in \mathbb{R} \, \, (\alpha, \beta=1,\cdots, n).
%v_i^\alpha \qquad (t \rightarrow -\infty) .
%V_i^\alpha(t) \simeq  \exp(\vert \kappa^\alpha \vert \, t ) \, v_i^\alpha \qquad (t \rightarrow -\infty) .
\end{eqnarray*}
Without loss of generality, one can set 
${\rm e}^{-\kappa^\beta t_0}  C_0^{\beta \alpha} = \delta^{\beta \alpha}$.
}
\begin{equation}
\label{eq:downward-flow-equation-tangent-vectors}
\frac{d}{dt} V_{z i}^\alpha(t) = \bar \partial_i   \bar \partial_j  \bar S[ \bar z]  \, \, \bar V^\alpha_{z j}(t)\qquad 
(\alpha = 1, \cdots, n).
\end{equation}
Indeed, $g \equiv \bar \partial \bar S[\bar z]$ itself satisfies this flow equation 
and it is expanded in terms of 
$\{ V_z^\alpha \}$ as 
$ g = \bar \partial \bar S[\bar z] = V_z^\alpha g^\alpha$ with $n$ real constants  $g^\alpha \in \mathbb{R} (\alpha=1,\cdots, n)$. 
It also follows that $\{ V_z^\alpha \}$ satisfy a reality condition,\footnote{
To show the reality condition, one should note 
\begin{equation*}
\frac{d}{dt} {\rm Im}\{  \bar{V}_z^\alpha(t)  V_z^\beta(t) \}
=  {\rm Im}\{ V_z^\alpha \partial^2 S[z] V_z^\beta(t) + \bar V_z^\alpha \bar \partial^2 \bar S[\bar z] \bar V_z^\beta(t) \} = 0, 
\end{equation*}
and 
\begin{equation*}
 {\rm Im}\{  \bar{V}_z^\alpha(t)  V_z^\beta(t) \} =  {\rm Im}\{  \bar{v}^\alpha v^\beta \} \exp(\kappa^\alpha t) \exp(\kappa^\beta t) =0   \qquad (t \ll 0).
\end{equation*}
}
\begin{equation}
\label{eq:tangent-vectors-reality-condition}
%i \, {\rm Im}\big\{ \bar{V}_{z i}^\alpha  V_{z i}^\beta \big\} =  
\bar{V}_{z i}^\alpha  V_{z i}^\beta - \bar{V}_{z i}^\beta V_{z i}^\alpha =0 \qquad 
(\alpha, \beta = 1, \cdots, n). 
\end{equation}
The basis of tangent vectors $\{ V_z^\alpha \}$, which satisfy the flow equations eq.~(\ref{eq:downward-flow-equation-tangent-vectors}),  is not orthonormal in general. One can make it orthonormal by  
Gram-Schmidt orthonormalization, or Iwasawa decomposition. 
In fact, $\{V_z^\alpha\}$ can be expressed in the following form, 
\begin{equation}
V_{z}^\alpha = U_{z}^\beta \, E^{ \beta\alpha} . 
\end{equation}
where $\{ U_z^\alpha \}$ is a orthonormal basis and $E$ is a real upper triangle matrix.\footnote{
By the 
Iwasawa decomposition, 
$V_z$ can be expressed in the form
$V_z= U_z D N$ , 
where $U_z$ is unitary, $D$ is positive diagonal, and $N$ is  upper  triangle with the unit diagonal elements. But, from the property $\bar{V}_{zi}^\alpha V_{zi}^\beta= \bar{V}_{zi}^\beta V_{zi}^\alpha$, one can show further 
that $N$ is real. 
Therefore, there exists  a real upper triangle matrix $E=DN$,  
and the tangent vectors $\{ V_z^\alpha \}$ are related to the orthonormal tangent vectors $\{ U_z^\alpha \}$ by
$V_z^\alpha = U_z^\beta \, E^{ \beta\alpha}$. 
}
In the vicinity of $z$, therefore, 
the thimble can be parametrized by real orthogonal coordinates  
$\{ \delta \xi^\alpha \} (\alpha =1,\cdots, n)$
such that $\delta z =   U_{z}^\alpha \delta \xi^\alpha$, $ \vert \delta z \vert^2 = \delta \xi^2$, and 
$\left. d^n z \, \right\vert_{ {\cal J}_\sigma} = d^n \delta \xi  \, \det U_z $.
Thus the measure on the thimbles, 
$\left. {\cal D}[z] = d^n z \right\vert_{ {\cal J}_\sigma}$, gives rise to an extra complex phase defined by 
\begin{equation}
{\rm e}^{i \phi_z} 
= \det U_z
=\frac{\phantom{\vert}\det V_z \phantom{\vert}}{\vert \det V_z \vert}.
\end{equation}

Given the tangent space $T_z$ and
the basis of  tangent vectors $\{ V_z^\alpha \} (\alpha =1,\cdots, n)$, 
directions normal to the thimble  at $z \in \mathcal{J}_\sigma$ 
are determined by the set of normal vectors 
$\{  i U_z^\alpha \}$ or 
$\{  i V_z^\alpha \}$$(\alpha =1,\cdots, n)$. This is because the reality condition 
eq.~(\ref{eq:tangent-vectors-reality-condition}) implies that 
\begin{equation}
{\rm Re}\big\{ (- i) \bar{V}_{z i}^\alpha \,  V_{z i}^\beta  \big\} = 0 \qquad 
(\alpha, \beta = 1, \cdots, n), 
\end{equation}
and $\{ i V_z^\alpha \}$ are orthogonal 
to $\{ V_z^\beta \}$ with respect to the inner product in $\mathbb{R}^{2n}$.

%Since a thimble ${\mathcal J}_\sigma$ is the union of all downward flows 
%which trace back to the critical point $z_\sigma$ at $t=-\infty$, 
%one may identify 
Finally, any point  $z$ on the thimble ${\mathcal J}_\sigma$ is identified uniquely  
by the {\em direction of the flow} on which $z$ lies and the {\em time of the flow} to get to $z$,  
both defined referring to a certain asymptotic region close to the critical point. 
In fact, the asymptotic solutions to the flow equations eqs.~(\ref{eq:downward-flow-equation}) and (\ref{eq:downward-flow-equation-tangent-vectors}) for $t \ll 0$ 
% in the limit $t \rightarrow -\infty$ 
%for a large negative time $t \ll 0$ 
can be expressed without loss of generality by
\begin{eqnarray}
\label{eq:flow-in-asymptotic-region}
 && z(t) \, \, \, \, \, \simeq  \, z_\sigma + v^\alpha  \exp( \kappa^\alpha  t) \,  e^\alpha \, ; \qquad 
e^\alpha e^\alpha = n, \\
\label{eq:tangent-vectors-flow-in-asymptotic-region}
&& V_z^\alpha(t) \simeq \, v^\alpha  \exp(\kappa^\alpha \, t ) ,  
\end{eqnarray}
and one can define the {\em direction of the flow}  by %the normalized vector 
$e^\alpha$ $(\alpha=1,\cdots,n; \| e \|^2 = n)$
and the {\em time of the flow} by $t^\prime=t-t_0$ 
with a certain reference time $t_0\ll 0$.\footnote{$t_0$ should be chosen so that $\| \epsilon \|^2 \ll n$ 
where $\epsilon^\alpha \equiv \exp(\kappa^\alpha t_0) e^\alpha$ and 
the linear approximation of the flow equation is valid.}
One can then define a map 
$z[e,t^\prime] : (e^\alpha,  t^\prime)  \rightarrow z \in {\mathcal J}_\sigma$ by 
\begin{eqnarray}
&& z[e, t^\prime] \, \, \, \, \, = \, z(t) \vert_{t=t^\prime+t_0}, 
%, \\
%&& V^\alpha[e, t^\prime]  = \, V^\alpha(t) \vert_{t=t^\prime+t_0}, 
\end{eqnarray}
provided the asymptotic form of the flow $z(t)$ is given 
by eq.~(\ref{eq:flow-in-asymptotic-region}).\footnote{In \cite{Mukherjee:2013aga}, 
a similar map between a thimble and its asymptotic ``Gaussian" region
has been introduced.}
% using the direction vector $e^\alpha$ and the scale parameter $t^\prime$.}
%, 
%and one can denote the fact $z \in {\mathcal J}_\sigma$ by $z = z[e, t^\prime]$. 
Moreover, under infinitesimal variations of the parameters $(e^\alpha, t^\prime)$,  
the variation of $z[e, t^\prime]$ is given by the following formula, 
\begin{eqnarray}
\delta z[e, t^\prime]
%&=& \delta n^\alpha V^\alpha[ z, \bar z] + \delta \tau \bar \partial \bar S[\bar z ] \\
&=& V_z^\alpha[e, t^\prime] \, %\exp(\vert \kappa^\alpha \vert \, t_0 ) \,
 ( \delta e^\alpha + \kappa^\alpha  e^\alpha \delta t^\prime  ) .
\end{eqnarray}
This is because an infinitesimal variation of the flow $\delta z(t)$ itself satisfies the flow equation for a tangent vector, 
\begin{equation}
\delta \dot{z}_i(t) = \bar \partial_i \bar \partial_j \bar S[ \bar z] \, \delta z_j(t) ,
\end{equation}
and it should be expanded in terms of 
$\{ V_z^\alpha \}$ as 
$\delta z(t) = V_z^\alpha(t) \, \delta c^\alpha$ with constants  $\delta c^\alpha \in \mathbb{R} \, (\alpha=1,\cdots, n)$. 
These constants $\{ \delta c^\alpha\}$ may be determined from 
the asymptotic form of $\delta z(t)$ for $t \ll 0$, %in the limit $t \rightarrow - \infty$,  
\begin{eqnarray}
\delta z(t) &=& v^\alpha \, \exp(\kappa^\alpha \, t ) 
(\delta e^\alpha + e^\alpha \kappa^\alpha \delta t ) \\
&=& V_z^\alpha(t) (\delta e^\alpha +  \kappa^\alpha e^\alpha \delta t ) 
\qquad \qquad \qquad \qquad (t \ll 0), 
\end{eqnarray}
and one obtains 
%$\delta z_i (t) = V_i^\alpha(t) (\delta e^\alpha +  \kappa^\alpha e^\alpha \delta t )$ and 
$\delta c^\alpha = \delta e^\alpha + \kappa^\alpha e^\alpha \delta t$.\footnote{In a similar reasoning, one obtains  $g^\alpha = \kappa^\alpha e^\alpha$, i.e. 
$ g= \bar \partial \bar S[\bar z] = V_z^\alpha \, \kappa^\alpha e^\alpha$. 
}

\section{An algorithm of hybrid Monte Carlo on Lefschetz thimbles}
\label{sec:HMC-on-Lefschetz-thimble}
%In this section, we next 
Next we describe a hybrid Monte Carlo algorithm which is applicable to the lattice models 
defined on Lefschetz thimbles.
Because of saddle-point structures of Lefschetz thimbles, 
it is not easy to keep track of a thimble
in stochastic processes like Langevin and hybrid Monte Carlo updates. 
%The deviation from the thimble easily grows 
%and the field configuration falls into the region where the real part 
%of the action is not bounded from below.
It is then necessary to be able to locate a field configuration 
on the thimble, $z \in \mathcal{J}_\sigma$, precisely in $\mathbb{C}^n$ and 
to constrain the field configuration onto the thimble in every stochastic step.
We therefore parametrize any point on the thimble $z \in \mathcal{J}_\sigma$ uniquely by 
the {\em direction of the flow}  $e^\alpha$$(\alpha=1,\cdots,n; \| e \|^2 = n)$
and the {\em time of the flow}  $t^\prime=t-t_0$ 
with a fixed reference time $t_0 (\ll 0)$, as discussed in the previous section.
Moreover,  regarding the molecular dynamics in hybrid Monte Carlo, 
we consider a constrained dynamical system including the forces normal to the thimble, i.e. 
along the normal vectors $\{  i V_z^\alpha \}$$(\alpha =1,\cdots, n)$, with
Lagrange multipliers. To integrate the equations of motion of the constrained system, 
we employ a second order constraint-preserving symmetric %(simplectic) 
integrator\cite{Leimkuhler-Reich}. 
%, which are 
%normal to the thimble. 

\subsection{To generate a thimble by solving the flow equations downward}
%and one can denote the fact $z \in {\mathcal J}_\sigma$ by $z = z[e, t^\prime]$. 
For given parameters $(e^\alpha, t')$ and $t_0 (\ll 0)$,  we generate the point $z[e, t^\prime] \in {\mathcal J}_\sigma$ by solving the flow equations 
eqs.~(\ref{eq:downward-flow-equation}) and (\ref{eq:downward-flow-equation-tangent-vectors}) 
downward with the initial conditions, 
\begin{eqnarray}
\label{eq:initial-condition-z}
z_i(t_0) &=& {z_\sigma}_i + v_i^\alpha \, \exp(  \kappa^\alpha  t_0) \, e^\alpha, \\
%\qquad [ \epsilon^\alpha \equiv n^\alpha \exp(  \kappa^\alpha  t_0) ]\\
\label{eq:initial-condition-V}
V_{z i}^\alpha(t_0) &=& v_i^\alpha \, \exp(  \kappa^\alpha  t_0).
\end{eqnarray}
We employ the fourth-order Runge-Kutta method with 
the number of iterations $n_{\rm lefs}$
and the size of  increment $h \equiv t'/n_{\rm lefs}$.
%\footnote{Namely, we repeat the following steps by 
%the number of iterations $n_{\rm lefs}$
%with the increment $h \equiv t'/n_{\rm lefs}$:
%\begin{eqnarray*}
%f_1 &=& \bar \partial \bar S[ z[t] ] \\
%f_2 &=& \bar \partial \bar S[ z[t] + (h/2) f_1 ] \\
%f_3 &=& \bar \partial \bar S[ z[t] + (h/2) f_2 ] \\
%f_4&=&  \bar \partial \bar S[ z[t] + h f_3 ] \\
%z(t+h) &=& z(t)+\frac{h}{6} \left( f_1 + 2 f_2 + 2 f_3 + f_4 \right) 
%\end{eqnarray*}
%\begin{eqnarray*}
%F^\alpha_1 &=& \bar V^\alpha(t) \bar \partial \bar \partial \bar S[ z[t] ] \\
%F^\alpha_2 &=& \left\{ \bar V^\alpha(t) +(h/2) \bar F^\alpha_1 \right\} \bar \partial\bar \partial \bar S[ z[t] + (h/2) f_1 ] \\
%F^\alpha_3 &=& \left\{ \bar V^\alpha(t) +(h/2) \bar F^\alpha_2 \right\}  \bar \partial\bar \partial \bar S[ z[t] + (h/2) f_2 ] \\
%F^\alpha_4&=&   \left\{ \bar V^\alpha(t)+\qquad h \bar F^\alpha_3 \right\}  \bar \partial\bar \partial \bar S[ z[t] + h f_3 ] \\
%V^\alpha(t+h) &=& V^\alpha(t)+\frac{h}{6} \left( F^\alpha_1 + 2 F^\alpha_2 + 2 F^\alpha_3 + F^\alpha_4 \right) 
%\end{eqnarray*}
%}
\footnote{
The computation of the tangent vectors $\{ V_z^\alpha \}$ $(\alpha =1,\cdots, n)$ is numerically very demanding.  We have used  GPUs in executing this computation.
}

To verify the solutions, one may check if the following relation is satisfied:
\begin{equation}
\bar \partial_i \bar S \big[\bar z[e, t^\prime] \big] 
-V_{z i}^\alpha [e, t^\prime]  \, \kappa^\alpha  e^\alpha = 0. 
\end{equation}
In what precision this relation holds would depend on several conditions and parameters.
First of all, it depends on the sizes of 
$\| z(t_0)- z_\sigma \|$ and ${\rm Re}\big( S[z(t_0)]-S[z_\sigma] \big)$,
which indicate how close to the critical point $z_\sigma$ the reference point $z(t_0)$ is. 
It depends also on 
the parameters of the Runge-Kutta method, $n_{\rm lefs}$ and  $h \equiv t'/n_{\rm lefs}$, 
and the size of the system, $n$. 

%We also compute 
Once the matrix $V_z=( V_{z i}^\alpha )$ is obtained, 
 its inverse $V_z^{-1}=( \{ V_z^{-1} \}^\alpha_i )$
such that 
$\sum_{\beta} V^\beta_{zi} \{ V_z^{-1} \}^\beta_j = \delta_{ij}$
and its determinant $\det V_z$ 
are computed through LU decomposition.  

\subsection{Constrained molecular dynamics}

%In order to
To formulate the molecular dynamics on the thimble $\mathcal{J}_\sigma$, we introduce 
a dynamical system defined by the equations of motion,\footnote{We 
use the abbreviation, $\frac{d}{d \tau} y(\tau) = \dot{y}$, where $\tau$ denotes 
the time coordinate of the dynamical system.}
\begin{eqnarray}
\label{eq:equations-of-motion-with-Lagrange-multipliers-z}
\dot z_i &=& w_i ,  \\
\label{eq:equations-of-motion-with-Lagrange-multipliers-w}
\dot w_i &=& - \bar \partial_i \bar S[\bar z] - i V_{z i}^\alpha \, \lambda^\alpha, 
\end{eqnarray}
and the constraints,
\begin{eqnarray}
\label{eq:eq-motion-constraints-z}
z_i &=& z_i[e, t^\prime], % \\
%w_i &=& V_{z i}^\alpha[e, t^\prime] \, w^\alpha , \quad w^\alpha \in \mathbb{R}   \qquad \text{or} \quad 
%{\rm Im}\left[  \{V_z^{-1}\}^\alpha_j w_j \right] = 0,
\end{eqnarray}
where $w_i$ are the momenta conjugate to $z_i$ and 
$\lambda^\alpha \in \mathbb{R}$ $(\alpha=1, \cdots, n)$ are the Lagrange multipliers.\footnote{
%, 
%and the matrix 
%$V_z^{-1}=( \{ V_z^{-1} \}^\alpha_i )$ is the inverse of 
%$V_z=( V_{z i}^\alpha )$ such that 
%$\sum_{\beta} V^\beta_{zi} \{ V_z^{-1} \}^\beta_j = \delta_{ij}$
The molecular dynamics on Lefschetz thimbles may be formulated by a Hamilton system on 
Riemann manifolds\cite{Girolami-Calderhead}. 
For example,  one may introduce auxiliary dynamical variables 
$x^\alpha \equiv \exp(\kappa^\alpha (t^\prime+ t_0)) \, e^\alpha$ 
and 
%conjugate momenta $p^\alpha = G^{\alpha \beta}[x] \, \dot{x}^\beta$, where 
the metric %$G^{\alpha \beta}[x]$ is given by 
$G^{\alpha \beta}[x] \equiv V_{z i}^\alpha[e,t^\prime] \bar V_{z i}^\beta[e,t^\prime] 
\, \exp(-\kappa^\alpha (t^\prime+t_0))  \, \exp(-\kappa^\beta (t^\prime+t_0))$ so that
$\vert\vert \delta z \vert\vert^2 
%= V_i^\alpha[e,t^\prime] \bar V_i^\beta[e,t^\prime]
%( \delta e^\alpha +  \kappa^\alpha  e^\alpha \delta t^\prime )
%( \delta e^\beta + \kappa^\beta e^\beta \delta t^\prime )
=G^{\alpha \beta}[x] \delta {x}^\alpha \delta {x}^\beta$.
One may then consider the Hamilton system with a {\em non-separable} Hamiltonian, 
\begin{equation}
H = \frac{1}{2} \{G^{-1}\}^{\alpha \beta}[x] \, p^\alpha p^\beta 
+ \frac{1}{2} \left\{ S + \bar S\right\}[x]  
+  \frac{1}{2} {\rm Tr Ln}( G[x] ).
\end{equation}
The equations of motion of this system may be solved by an {\em implicit}
second order symplectic integrator. 
}
It  follows from the equations of motion eqs.~(\ref{eq:equations-of-motion-with-Lagrange-multipliers-z}), (\ref{eq:equations-of-motion-with-Lagrange-multipliers-w}) 
and the constraint eq.~(\ref{eq:eq-motion-constraints-z}) that
\begin{eqnarray}
%z_i &=& z_i[e, t^\prime],  \\
\label{eq:eq-motion-constraints-w}
w_i &=& V_{z i}^\alpha[e, t^\prime] 
\, w^\alpha , \quad w^\alpha \in \mathbb{R}   \qquad \text{or} \quad 
{\rm Im}\left[  \{V_z^{-1}\}^\alpha_j w_j \right] = 0. 
\end{eqnarray}
In this system, a  conserved Hamiltonian is given by 
\begin{equation}
H = \frac{1}{2} \bar{w}_i w_i  + \frac{1}{2} \left\{ S[z ] + \bar S [ \bar z] \right\}.
\end{equation}
It follows indeed that
\begin{eqnarray}
\dot{H} 
&=& \frac{1}{2} \{ \dot{\bar{w}}_i w_i + \bar{w}_i \dot{w}_i \} 
 +\frac{1}{2} \left\{ \partial_i S[z ] \dot{z}_i + \bar \partial_i \bar S [ \bar z] \dot{\bar{z}}_i \right\}  \nonumber\\
 &=&  \frac{1}{2}\left\{   ( + i \bar{V}_{z i}^\alpha \lambda^\alpha ) w_i + \bar{w}_i  (- i {V}_{z i}^\alpha \lambda^\alpha)\right\} \nonumber\\
 &=& \frac{i}{2} \lambda^\alpha w^\beta \left\{ \bar{V}_{z i}^\alpha {V}_{z i}^\beta - \bar{V}_{z i}^\beta {V}_{z i}^\alpha \right\} =0.
 \end{eqnarray}

\begin{figure}[htbp]
\begin{center}
\includegraphics[width=9cm]{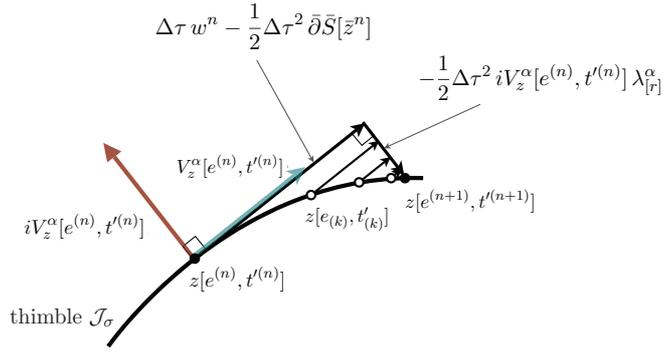}
\end{center}
\caption{A fixed-point method to solve the constraint eq.~(\ref{eq:constraint-z}).}
\label{fig:how-to-solve-constraint}
\end{figure}

To integrate the equations of motion with the Lagrange 
multipliers eqs.~(\ref{eq:equations-of-motion-with-Lagrange-multipliers-z}) and (\ref{eq:equations-of-motion-with-Lagrange-multipliers-w}), we employ 
the second order constraint-preserving symmetric integrator\cite{Leimkuhler-Reich}:
it is  assumed first that $z^n$ and $w^n$ satisfy the constraints
\begin{eqnarray}
z^n &=& z[{e}^{(n)},t^{\prime (n)}] , \\
w^n &=& V_z^\alpha[{e}^{(n)},t^{\prime (n)}] \, {w^\alpha}^{(n)},   \quad {w^\alpha}^{(n)} \in \mathbb{R}, 
%\qquad \text{or} \quad 
%{\rm Im}\left[  \{V_z^{-1}[{e}^{(n)},t^{\prime (n)}] \}^\alpha_j w^n_j \right] = 0,
\end{eqnarray}
and $z^{n+1}$ and $w^{n+1}$ are then determined for a given step size $\Delta \tau$ by 
\begin{eqnarray}
\label{eq:constraint-preserving-integrator-1}
{w}^{n+1/2} &=& {w}^{n} \qquad  - \frac{1}{2} \Delta \tau  \, \bar \partial \bar S[\bar z^n] \, \, \, \, - \frac{1}{2} \Delta \tau \,   i V_z^\alpha[{e}^{(n)},t^{\prime (n)}]  \, \lambda_{[r]}^\alpha,  \\
\label{eq:constraint-preserving-integrator-2}
z^{n+1} &=& z^n \qquad  + \Delta \tau \, {w}^{n+1/2}, \\
\label{eq:constraint-preserving-integrator-3}
{w}^{n+1} &=& {w}^{n+1/2} - \frac{1}{2} \Delta \tau  \, \bar \partial \bar S[\bar z^{n+1}] - \frac{1}{2} \Delta \tau \,  i V_z^\alpha[{e}^{(n+1)},t^{\prime (n+1)}] \, \lambda_{[v]}^\alpha,  
\end{eqnarray}
where $\lambda_{[r]}^\alpha$ and $\lambda_{[v]}^\alpha$ are fixed by imposing the constraints,
\begin{eqnarray}
\label{eq:constraint-z}
%\lambda_{[r]}^\alpha \quad  \text{fixed by imposing } 
&& z^{n+1} = z[{e}^{(n+1)},t^{\prime(n+1)}],  \\
\label{eq:constraint-w}
%\lambda_{[v]}^\alpha \quad \text{fixed by imposing } 
&&
{w}^{n+1} 
= V_z^\alpha[{e}^{(n+1)},t^{\prime(n+1)}] \, {w^\alpha}^{(n+1)} ,   
\quad {w^\alpha}^{(n+1)} \in \mathbb{R},  
\end{eqnarray}
respectively.
The first constraint eq.~(\ref{eq:constraint-z}) reads
\begin{eqnarray}
\label{eq:constraint-z-2}
%z[{n^\alpha}^{(n)} + \Delta n^\alpha , \tau^{(n)}+\Delta \tau] - z[{n^\alpha}^{(n)} , \tau^{(n)}]
%= \Delta t \, {w}^{n+1/2}
z[{e}^{(n+1)}, t^{\prime (n+1)}] - z[{e}^{(n)} , t^{\prime (n)}]
&=&   \Delta \tau \, {w}^{n}  - \frac{1}{2} \Delta \tau^2  \, \bar \partial \bar S[\bar z^n]  \nonumber\\
&&\qquad \qquad   - \frac{1}{2} \Delta \tau ^2\, i V_z^\alpha[{e}^{(n)},t^{\prime (n)}] \, \lambda_{[r]}^\alpha.  
\end{eqnarray}
This is solved by a {\em fixed-point} iteration method\footnote{This method to find 
$({e^\alpha}^{(n+1)}, t^{\prime (n+1)} )$ and $\lambda_{[r]}^\alpha$ in eq.~(\ref{eq:constraint-z-2})
can
 also be used in Langevin-type updates.}: 
to find $({e^\alpha}^{(n+1)}, t^{\prime (n+1)} )$ and $\lambda_{[r]}^\alpha$, 
%Seek $\Delta n^\alpha$ and $\Delta \tau$ iteratively so that
we generate the sequences 
$( e^\alpha_{(k)}, t^\prime_{(k)}) \, (k=0,1,\cdots)$ with 
$( e^\alpha_{(0)}, t^\prime_{(0)})=({e^\alpha}^{(n)} , t^{\prime (n)} )$
and ${ \lambda_{[r]}^\alpha  }_{(k)} \, (k=0,1,\cdots)$
%by assuming that 
so that 
the increments,  
\begin{eqnarray}
\Delta e^\alpha_{(k)} &=& e^\alpha_{(k+1)} - e^\alpha_{(k)}, \qquad  
\sum_{\alpha=1}^n \Delta e^\alpha_{(k)} e^{\alpha (n)} =0, \\
\Delta t^\prime_{(k)} &=& t^\prime_{(k+1)} - t^\prime_{(k)}, 
\end{eqnarray}
are infinitesimal and  
$( \Delta {e^\alpha}_{(k)}, \Delta t^\prime_{(k)})$ and ${ \lambda_{[r]}^\alpha  }_{(k)}$ 
are determined by 
\begin{eqnarray}
  \Delta {e^\alpha}_{(k)} + e^{\alpha (n)} \kappa^\alpha \Delta t^\prime_{(k)} 
 &=& {\rm Re} \left[ \{ V_z^{-1}[{e}^{(n)} , t^{\prime(n)}] \}^\alpha_i  \times \phantom{\frac{1}{2}} \right. 
 \nonumber\\
 && \left. \qquad %\quad 
 \big( 
 z_i [{e}^{(n)} , t^{\prime(n)}]  
 +
 \Delta \tau \, {w}^{n}_i  - \frac{1}{2} \Delta \tau ^2  \, \bar \partial_i \bar S[\bar z^n]
 - z_i [{e}_{(k)} , t^\prime_{(k)}] 
 \big)
 \right], 
 \nonumber\\
 && \\
 \frac{1}{2} \Delta \tau^2\, { \lambda_{[r]}^\alpha  }_{(k)}
  &=& 
{\rm Im} \left[ \{ V_z^{-1} [{e}^{(n)} , t^{\prime(n)}]\}^\alpha_i
\big( z_i [{e}^{(n)} , t^{\prime (n)}]  - z_i [{e}_{(k)} , t^\prime_{(k)}]\big) \right], 
\end{eqnarray}
until a stopping condition,  
\begin{equation}
\left\|  
V_z^\alpha[{e}^{(n)} , t^{\prime(n)}] 
\big(  \Delta {e^\alpha}_{(k)} + e^{\alpha (n)}  \kappa^\alpha \Delta t^\prime_{(k)} \big)
 \right\|^2
\le \, n \, {\epsilon'}^2 , 
\end{equation}
is satisfied for a sufficiently small $\epsilon'$ to achieve a given precision.\footnote{The squared 
norm  of $e^\alpha_{(k+1)}$ has the second order correction, 
$\| e^\alpha_{(k+1)} \|^2= \|  e_{(k)}+ \Delta e_{(k)} \|^2 = n+ (\Delta e_{(k)})^2$,  and  
it is renormalized 
%by the factor $1/\sqrt{1+(\Delta e_{(k)})^2/n }$.
as $e^\alpha_{(k+1)} \rightarrow e^\alpha_{(k+1)}/\sqrt{1+(\Delta e_{(k)})^2/n }$.}
(See fig.~\ref{fig:how-to-solve-constraint}.) 
Once $(e^{\alpha (n+1)}, t^{\prime (n+1)} )$ and $z[e^{(n+1)}, t^{\prime (n+1)}]$
are obtained, we compute the set of tangent vectors $\{ V^\alpha_z[e^{(n+1)}, t^{\prime (n+1)} ]\}$ and the inverse matrix $V_z^{-1}[e^{(n+1)}, t^{\prime (n+1)} ]$.
The second constraint in eq.~(\ref{eq:constraint-w})
is then solved by
%In order to find $\lambda_{[v]}^\alpha$, solve 
\begin{equation}
\frac{1}{2} \Delta \tau \, { \lambda_{[v]}^\alpha  } = 
{\rm Im} \left[ \big\{ V_z^{-1}[e^{(n+1)},t^{\prime (n+1)}] \big\}^\alpha_i 
\big( {w}^{n+1/2}_i - \frac{1}{2} \Delta \tau  \, \bar \partial_i \bar S[\bar z^{n+1}]
\big) \right] . 
\end{equation}

\subsection{Hybrid Monte Carlo updates}

A hybrid Monte Carlo update then consists of the following steps 
for a given trajectory length $\tau_{\rm traj}$ and a number of steps $n_{\rm step}$:
%Compute  the initial ${\xi^\alpha}^{(0)}$ from $(\epsilon^\alpha, \tau)^{(0)}$.
\begin{enumerate}
\item Set the initial field configuration $z_i$:
\begin{equation}
\{ e^{\alpha (0)}, t^{\prime (0)} \} =\{ e^\alpha, t^\prime\}, \qquad z^0 = z[e ,t^{\prime} ]. 
\end{equation}

\item Refresh the momenta $w_i$ 
by generating $n$ pairs of unit gaussian random numbers $(\xi_i ,\eta_i)$, 
setting tentatively $w_i =\xi_i + i \eta_i$, and  chopping 
the non-tangential parts:

\begin{equation}
w^0= V_{z}^\alpha \, {\rm Re}[ \{V_z^{-1}\}^\alpha_j ( \xi_j+ i \eta_j) ] 
= U_z^\alpha \, {\rm Re}[ \{U_{z}^{-1}\}^\alpha_j ( \xi_j + i \eta_j)].
\end{equation}

\item Repeat $n_{\rm step}$ times of the second order symmetric integration eqs.~(\ref{eq:constraint-preserving-integrator-1})--(\ref{eq:constraint-w}) with the 
step size $\Delta \tau = \tau_{\rm traj} / n_{\rm step}$. 

\item Accept or reject by $\Delta H=H[w^{n_{\rm step}}, z^{n_{\rm step}}]-H[w^0,z^0]$.

\end{enumerate}

As for the initialization procedure, 
one may generate unit gaussian random numbers $\eta^\alpha$$(\alpha=1, \cdots, n)$,  set
\begin{equation}
e^{\alpha} =\eta^\alpha  \sqrt{\frac{n}{\sum_{b=1}^n \eta^b \eta^b}}, \qquad t^\prime = - t_0, 
\end{equation} 
and then prepare  $z[e, t^\prime]$, $\{ V_z^\alpha[e, t^\prime]\}$, 
and the inverse matrix $V_z^{-1}[e, t^\prime]$. 
%and 
% the residual phase factor 
%${\rm e}^{i \phi_z} = \det V_z / \vert \det V_z  \vert$. 

\subsection{To measure observables by reweighting the residual sign factors}

In the hybrid Monte Carlo method described above, 
the contribution of the residual phase factor, ${\rm e}^{i \phi_z} = \det V_z / \vert \det V_z  \vert$, is neglected.
To obtain the expectation value of an observable on the given thimble $\mathcal{J}_\sigma$, %$O[z]$, 
we need to evaluate  the average of the observable %$O[z]$ 
with the residual phase factor reweighed. 
%, ${\rm e}^{i \phi_z} = \det V_z / \vert \det V_z  \vert$, reweighed. 
Let us denote the simple statistical average of an operator  $o[z]$ on the thimble $\mathcal{J}_\sigma$
by $\langle o[z] \rangle^\prime_{\mathcal{J}_\sigma}$:
\begin{eqnarray}
\langle  o[z]  \rangle^\prime_{\mathcal{J}_\sigma} 
&=& \frac{1}{N_{\rm conf}}  \sum_{k=1}^{N_{\rm conf}} o[z^{(k)}],  %  {\rm e}^{i \phi_z}[z^{(k)}], 
\end{eqnarray}
where $N_{\rm conf}$ is the number of field configurations obtained by the hybrid Monte Carlo updates.
The expectation value of a given observable $O[z]$ on the thimble $\mathcal{J}_\sigma$
should then be evaluated by the following formula, 
\begin{equation}
\label{eq:observables-on-vacuum-thimble-residual-phase-quenched}
\left\langle O[z] \right\rangle_{\mathcal{J}_\sigma} = 
\frac
{\langle {\rm e}^{i \phi_z}  O[z] \rangle^\prime_{\mathcal{J}_{\sigma}}
}
{\langle {\rm e}^{i \phi_z}  \rangle^\prime_{\mathcal{J}_{\sigma}}
}. 
\end{equation}
For this formula eq.~(\ref{eq:observables-on-vacuum-thimble-residual-phase-quenched}) 
to work, it is crucial that
the averages of the residual sign factors, 
$\{ \langle {\rm e}^{i \phi_z}  \rangle^\prime_{\mathcal{J}_\sigma} \}$$(\sigma \in \Sigma)$, are not   
vanishingly small, in particular,  for the thimble associated with 
the classical vacuum, $\mathcal{J}_{\rm vac}$. 
This is the possible sign problem in our hybrid Monte Carlo method, which should  be studied carefully and systematically. 

%\newpage
\section{HMC simulations of the complexified $\lambda \phi^4$ model at finite density}
\label{sec:test}
%Finally we test the 
Now we test the hybrid Monte Carlo algorithm described in the previous section
by applying it 
to the complex $\lambda \phi^4$ model 
with chemical potential $\mu$\cite{Aarts:2008wh,Gattringer:2012df,Cristoforetti:2013wha}.
%, the so-called relativistic Bose gas  
%
The action of the model is defined in the lattice unit by 
\begin{eqnarray}
S 
&=& \sum_{ x \in \mathbb{L}^4 } \Big\{ 
 \big( \varphi^\dagger (x+\hat 0) {\rm e}^{+\mu} - \varphi^\dagger(x) \big)
 \big( {\rm e}^{-\mu} \varphi (x+\hat 0)  - \varphi(x) \big)
 \nonumber\\
 && \qquad
+ \sum_{k=1}^3 \vert \varphi (x+\hat k) - \varphi(x) \vert^2
+\frac{\kappa}{2} \varphi^\dagger (x) \varphi(x)
+\frac{\lambda}{4}  \big(  \varphi^\dagger (x) \varphi(x)\big)^2
         \Big\} 
%         &+& \sum_{ x \in \mathbb{L}^n } h \, \phi_1(x) 
\\
&=&
\sum_{ x \in \mathbb{L}^4 } \Big\{  
-\phi_a(x) \phi_b(x+\hat 0)  \big[\delta_{ab}  \cosh(\mu) - i \epsilon_{ab} \sinh(\mu) \big]
\nonumber\\
&& \qquad
-\sum_{k=1}^3 \phi_a(x) \phi_a(x+\hat k) 
+\frac{( 8 + \kappa)}{2} \phi_a(x) \phi_a(x)
+\frac{ \lambda}{4} \big(  \phi_a(x) \phi_a(x) \big)^2
         \Big\}, 
\end{eqnarray}
where $\varphi(x)=\big(\phi_1(x)+ i \phi_2(x)\big)/\sqrt{2}$
and the real field variables  $\phi_a(x) \in \mathbb{R}$ $(a=1,2)$ are used in the second expression.
We assume that  the lattice $\mathbb{L}^4$ is finite  
with a linear extent $L$ and a volume $V=L^4$,  %${\rm Vol}(\mathbb{L}^4)=$
and  the field variables satisfy the periodic boundary conditions.
In complexification, %and $h_0 = h \sqrt{K_0}$, 
the field variables are  complexified as $\phi_a(x) \rightarrow  z_a(x) \in \mathbb{C}$ $(a=1,2)$ and
rescaled for later convenience as $z_a(x) \rightarrow \sqrt{K_0} \, z_a(x)$ so that
$K_0 (8+\kappa) =1$ and $ K_0^2 \lambda = \lambda_0$. 
The complexified action then reads
\begin{eqnarray}
S[z]
&=& \sum_{ x \in \mathbb{L}^4 } 
\Big\{ 
+ \frac{1}{2} z_a(x) z_a(x)
+\frac{\lambda_0}{4}  \big( z_a(x) z_a(x) \big)^2 -K_0 \sum_{k=1}^3 z_a(x) z_a(x+\hat k) 
%+ h_0 \, \phi_1(x)
\nonumber\\
&& \qquad \quad
- K_0  \, z_a(x) z_b(x+\hat 0)  \big[\delta_{ab}  \cosh(\mu) - i \epsilon_{ab}  \sinh(\mu) \big]
\Big\} . 
\end{eqnarray}
%where $K_0= \frac{1}{(2D+\kappa)}$, $\lambda_0=K_0^2 \lambda$

Among possible critical points in this model, those with constant fields $z_a(x) = z_a$ are relatively easy to find.
%, and we consider these critical points. 
Such critical points are determined by the following stationary condition, 
\begin{equation}
\left. \frac{\partial S[z]}{\partial z_a(x)} \right\vert_{z_a(x) = z_a} 
= \left( 1- 6 K_0 - 2 K_0 \cosh(\mu) \right) z_a  + \lambda_0 ( z_1^2 + z_2^2 ) z_a  = 0  \quad (a=1,2). 
\end{equation}
There is a classical critical value in $\mu$,  for fixed $K_0 (< 1/8)$ and $\lambda_0 (>0)$,  given by 
\begin{eqnarray}
\tilde \mu_c &=& \ln \left[ 
\Big( \frac{1-6 K_0}{2K_0} \Big) + \sqrt{ \Big( \frac{1-6 K_0}{2K_0} \Big)^2 -1 }
\right] , 
%\\
% &=& 0.962 \ldots  \quad \text{for } K_0 = 1/(8 + \kappa) = 1/9 , 
\end{eqnarray}
and the solutions to the stationary condition are obtained as follows:
\begin{enumerate}
\item For $\mu \le \tilde \mu_c$, 
\begin{enumerate} 
\item  $z_1=z_2=0$ ; $S[z]= 0$, 
\item $z_1= i \phi_0 \cos \theta$,
         $z_2=i \phi_0 \sin \theta$ ; $S[z]=- L^4 \frac{\lambda_0}{4} \phi_0^4$, 
\\ where $\phi_0 = \sqrt{\frac{+\big(1-6 K_0 - 2 K_0 \cosh(\mu) \big)} {\lambda_0}}$.

\end{enumerate} 
\item For $\mu > \tilde \mu_c$, 
\begin{enumerate}
\item $z_1=z_2=0$ ; \quad $S[z]= 0$, 
\item $z_1= \phi_0 \cos \theta$,
         $z_2=\phi_0 \sin \theta$ ; $S[z]=- L^4 \frac{\lambda_0}{4} \phi_0^4$, 
\\ where $\phi_0 = \sqrt{\frac{-\big(1-6 K_0 - 2 K_0 \cosh(\mu) \big)} {\lambda_0}}$.

\end{enumerate}
\end{enumerate}
The solutions 1-(a), 2-(a), and 2-(b) are real.
They are in fact the classical solutions in the original model, and
the solutions 1-(a) and 2-(b) are the classical vacua for $\mu < \tilde \mu_c$ and $\mu > \tilde \mu_c$, respectively. 
The solution 1-(b) are pure imaginary, and 
the thimbles associated with this critical point do not contribute to
the path-integration, because $-  {\rm Re} S[z_\sigma] > \text{max}\left\{ - {\rm Re} S[x] \right\} (= 0 \text{ for } \mu < \tilde \mu_c)$.
% and it does not hold that 
%$-  {\rm Re} S[z_\sigma] >  \text{max}\left\{ - {\rm Re} S[x] \right\} (x \in {\mathcal C}_{\mathbb{R}}) = 0$.
In the solutions 1-(b) and 2-(b), the O(2) $\big( \text{U(1)} \big)$ symmetry breaks down spontaneously,  
and they give actually  the {\em critical regions} of real dimension one, parameterized by $\theta \in [0,2\pi]$.
%\footnote{For the original model with the real field variables, 
%it is known that if one applies the Hybrid Monte Carlo method with only the real part of the action ${\rm Re} S[x]$,  
%the average of the reweighting  complex phase factor $\exp(- i {\rm Im} S[x])$ vanishingly small in the 
%broken phase $\mu > \mu_c$}

We take the thimbles associated with 
the classical vacua, 1-(a) for $\mu < \tilde \mu_c$ and 2-(b) for $\mu > \tilde \mu_c$, 
for our purpose.
%\footnote{Close to the critical point $\tilde \mu_c$,  we expect that the contributions 
%of the other thimbles such as those associated with 1-(b) and 2-(a) become important.
%A study on this point would be reported in a forth coming paper. }
For the model parameters, 
we choose the values, $\kappa=1$ and $\lambda=1$,  
following the study in \cite{Aarts:2008wh}. In this case, $\tilde \mu_c \simeq 0.962$.  
We measure the number density,  
\begin{equation}
n[z] = \frac{1}{L^4} \sum_{x} K_0 \, z_a(x) z_b(x+\hat 0) \, \big[ \delta_{ab} \sinh(\mu) - i \epsilon_{ab} \cosh(\mu) \big]
\end{equation}
as well as the residual phase factor, ${\rm e}^{i \phi_z} = \det V_z / \vert \det V_z  \vert$,
for various values of $\mu$ in the range $\mu \in [0, 1.5]$.\footnote{In this model, the orthonormal tangent vectors at the critical point $\{ v_a(x)^\alpha \}$ $(\alpha=1,\cdots,2V)$ 
can be chosen to satisfy 
$C \bar v^\alpha = v^\beta P^{\beta \alpha}$,  
where $C$ is the charge conjuation operator defined by $C : z_1(x) \leftrightarrow z_2(x)$, while 
$P$ is a permutation operator. It then follows that 
${\rm e}^{i \phi_z}\vert_{z=z_{\rm vac}} =\det v =  \pm 1$.
%See appendix for the case of 2-(b) for $\mu > \tilde \mu_c$.
%In fact, $\bar S[\bar z ; \mu]=S[\bar z ; -\mu] =S[C\bar z; \mu]$ where $C$ is the charge conjuation operator defined by $C : z_1(x) \leftrightarrow z_2(x)$. This implies that 
%$\bar K[\bar z_\sigma; \mu] =K[\bar z_\sigma; - \mu]= C K[C \bar z_\sigma; \mu]C$ and 
%$( \det v )^2 = \prod_\alpha \kappa^\alpha / \det K \in \mathbb{R}$. ...
}
We consider only the lattice size $L=4$ in this work.

\subsection{Thimble 1-(a) for $\mu <  \tilde \mu_c$}

\begin{figure}[htbp]
\begin{center}
\includegraphics[width=7.5cm]{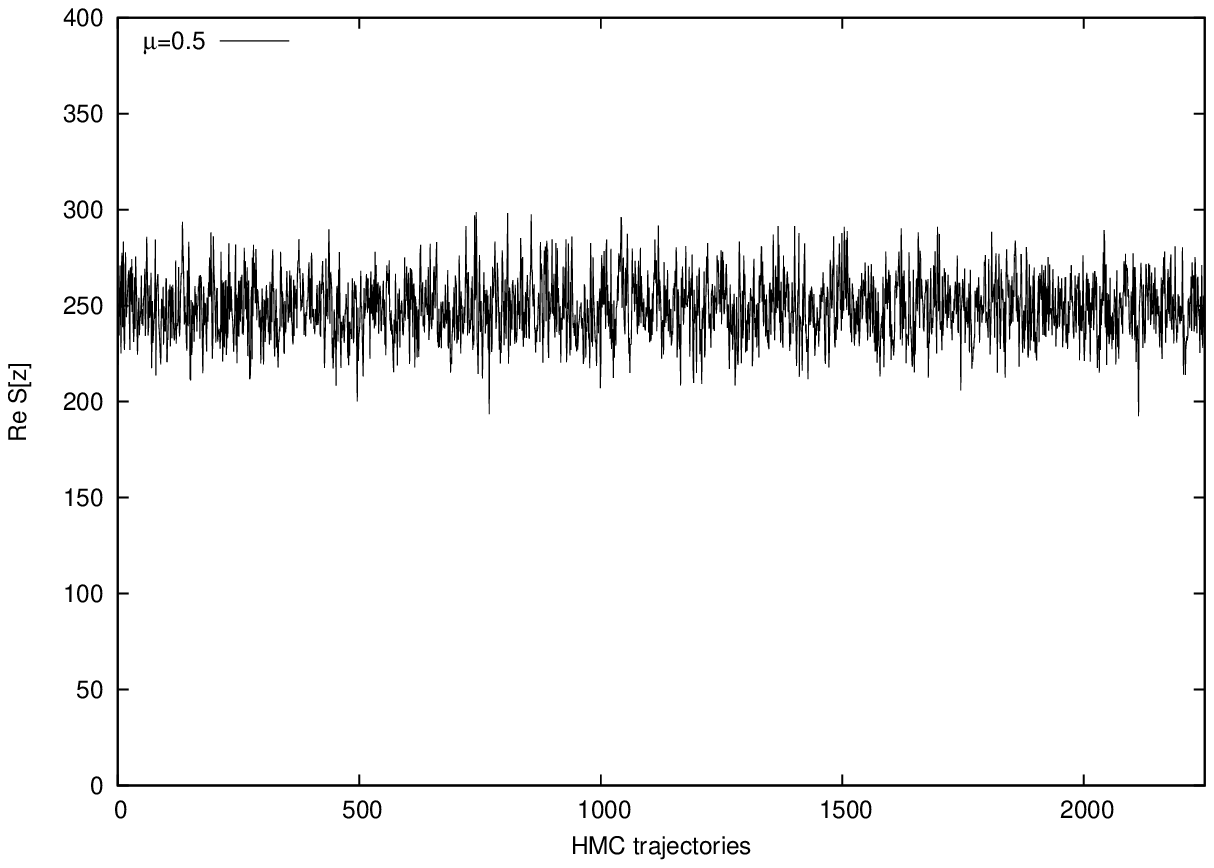}
\includegraphics[width=7.5cm]{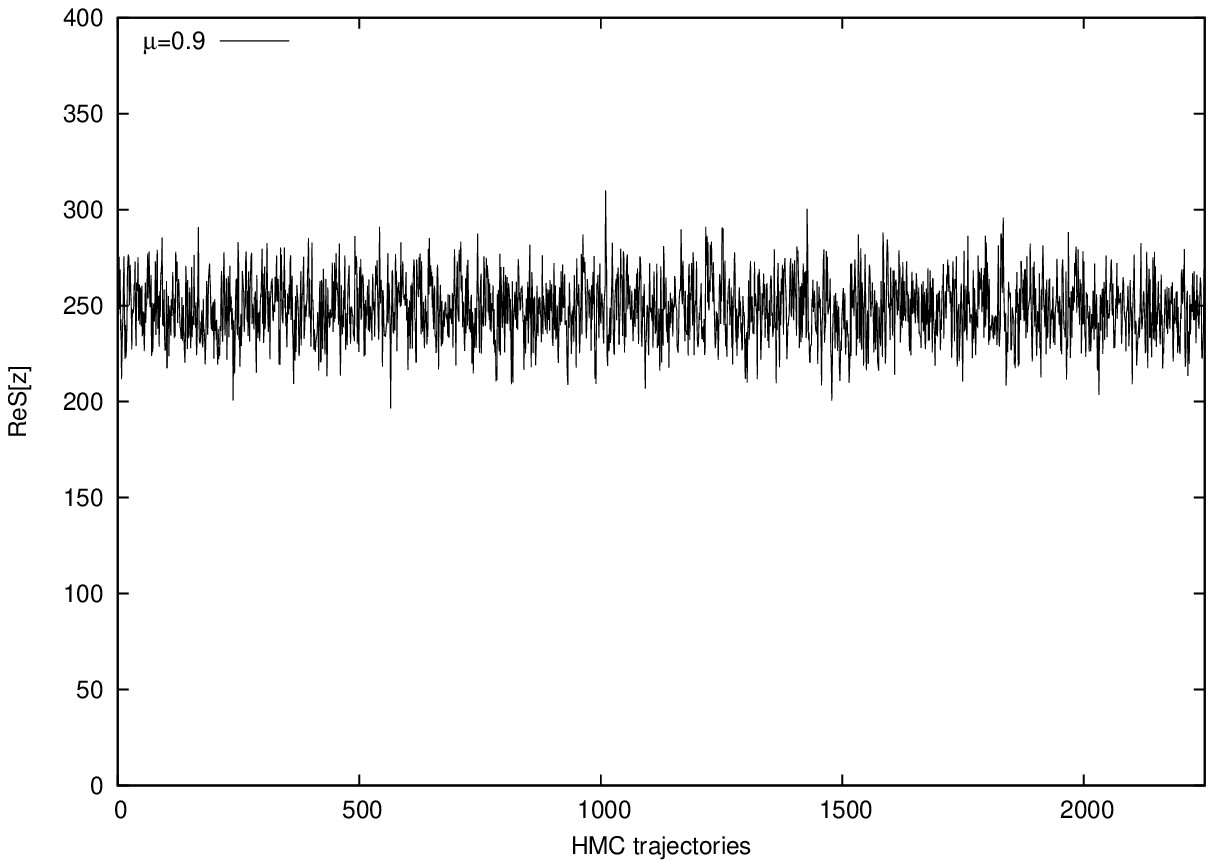}
\end{center}
\caption{Monte Carlo  histories of ${\rm Re} S[z] $ for $\mu=0.5$ and $0.9$ ($\kappa=1.0$, $\lambda=1.0$, $L=4$).
In the course of the MC updates,  the absolute values of ${\rm Im} S[z]$ were kept less than $1.0 \times 10^{-4}$.}
\label{figs:MC-histories-ReS}
\end{figure}

\begin{figure}[htbp]
\begin{center}
\vspace{2em}
\includegraphics[width=7.5cm]{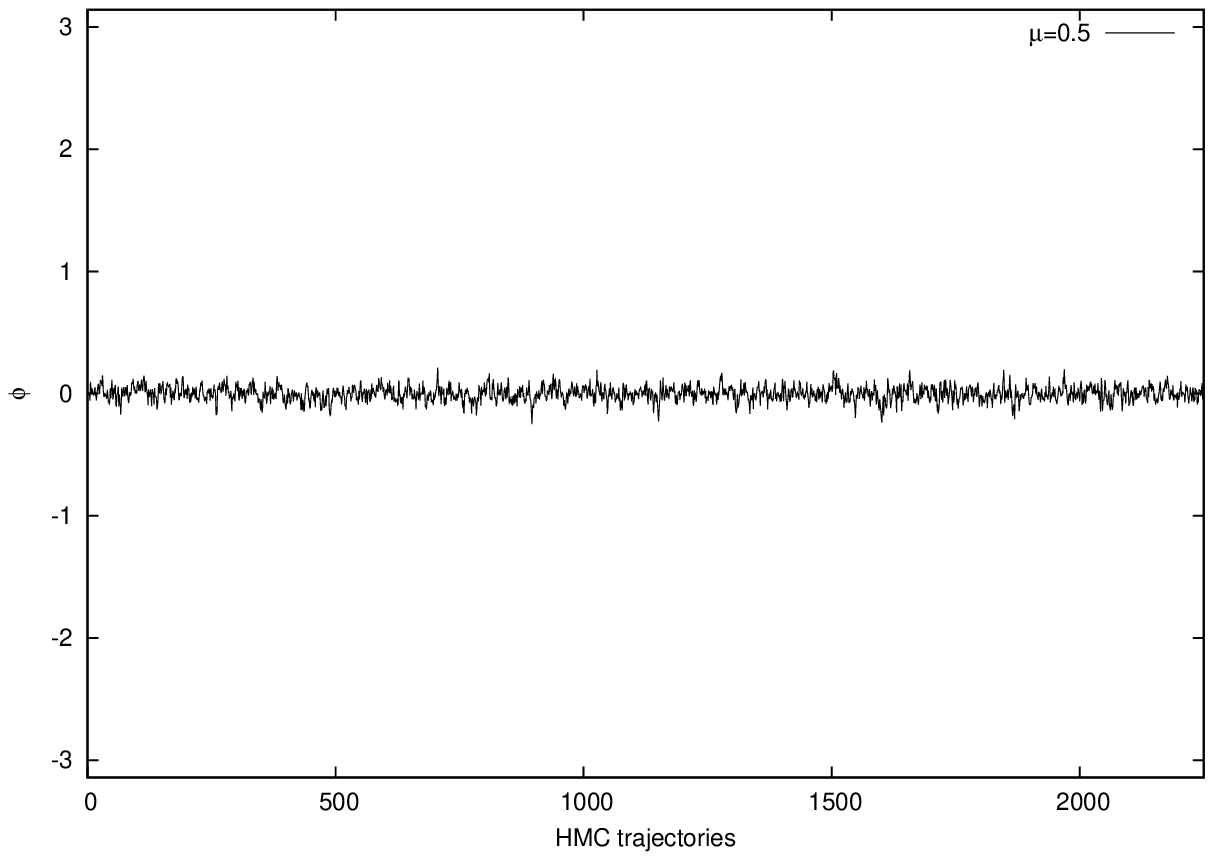}
\includegraphics[width=7.5cm]{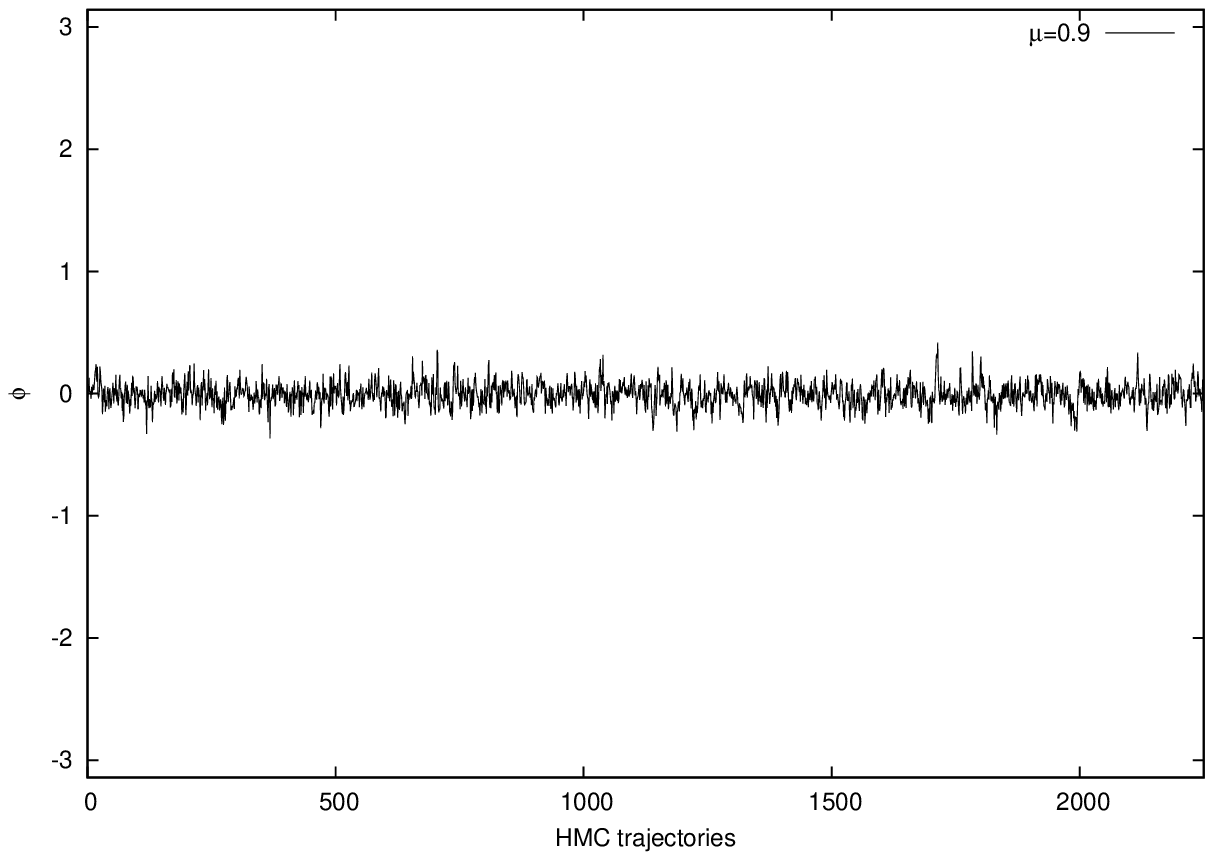}
\end{center}
\caption{Monte Carlo histories of $\phi_z$  
for $\mu=0.5$ and $0.9$  ($\kappa=1.0$, $\lambda=1.0$, $L=4$).}
\label{figs:MC-histories-phi}
\end{figure}

The algorithm given in section~\ref{sec:HMC-on-Lefschetz-thimble} 
applies straightforwardly to the thimble 1-(a) for $\mu <  \tilde \mu_c$.
We have generated $4,250$ trajectories  
for each value $\mu=0.1, 0.3, 0.5, 0.7$ and $0.9$ with the parameters listed 
in table~\ref{table:parameters-symmetric-phase}.
Each trajectory is of the length $\tau_{\rm traj}=1.0$ and  obtained in the number of 
steps $n_{\rm step}=20$. 
In solving the flow equations, the parameters are chosen as $t_0 =-5.0$ and $n_{\rm lefs}=100$.
%It was then found that %
We have found in the course of the simulations that
%$t^\prime \in [ 4.9, 5.1]$ 
the scale variable $t^\prime$ varies within the range $[ 4.9, 5.1]$
and $h=t^\prime/n_{\rm lefs} \simeq 0.05$ most of the time, 
and 
the solutions satisfy the bounds,  
$ \vert {\rm Im} S[z] \vert \lesssim 1.0 \times 10^{-4}$ and 
$\| \bar \partial \bar S - V^\alpha \kappa^\alpha e^\alpha \|^2 /2V \le 1.0 \times 10^{-4}$. 
In solving the constraint in the molecular dynamics, 
the fixed-point method converges with the iteration numbers $l \lesssim 4$
for the step size $\Delta \tau =\tau_{\rm traj}/ n_{\rm step}=0.05$ 
and the bound $\epsilon^\prime = 1.0 \times 10^{-3}$.
$\Delta H$ turns out to be rather small, and the acceptance rates are $\simeq$ 0.99 on average.     
The integrated auto-correlation times are estimated as $\tau_{\rm int} \simeq 2$ for ${\rm Re} S[z] $
and  $\tau_{\rm int} \simeq 3$ for $\phi_z$ for all the given values of $\mu$.
In fig.~\ref{figs:MC-histories-ReS}, 
Monte Carlo histories of %the real part of the action 
${\rm Re} S[z]$ are shown 
for $\mu=0.5$ and $0.9$. 
(As for %the imaginary part of the action 
${\rm Im} S[z]$, 
its absolute value is kept less than $1.0 \times 10^{-4}$ in all trajectories.)
In fig.~\ref{figs:MC-histories-phi}, Monte Carlo histories of the residual phase $\phi_z$ are shown for $\mu=0.5$ and $0.9$.

\begin{table}[htbp]
\caption{Simulation parameters for the thimble 1-(a) ($\mu < \tilde \mu_c$)}
\begin{center}
\begin{tabular}{| l | l | l |}
\hline
& Parameters & Resulting conditions \\ \hline
%Model & $\kappa=1.0$, $\lambda=1.0$, $L=4$ & \\ \hline
Thimble & $t_0 =-5.0$ &  $ \vert {\rm Re} \big( S[z(t_0)] - S[z_{\rm vac}] \big)\vert \lesssim 1.0 $ \\ 
(Solving flow eqs.) & $n_{\rm lefs}=100$ & $ \vert {\rm Im} S[z] \vert \lesssim 1.0 \times 10^{-4}$ \\
&$h=t^\prime/n_{\rm lefs} \simeq 0.05$& 
$\| \bar \partial \bar S - V^\alpha \kappa^\alpha e^\alpha \|^2 /2V \le 1.0 \times 10^{-4}$ \\ \hline
Molecular Dynamics  & $\tau_{\rm traj}=1.0$
                                  & scale variable range : $t^\prime \in [ 4.9, 5.1]$ \\ 
(Solving constraint)   & $n_{\rm step}=20$ & $\Delta H \lesssim 0.1$ \\ 
                                 &$\Delta \tau=0.05$ & acceptance rate $\simeq 0.99$\\
%                                 &&\\
&$\epsilon^\prime = 1.0 \times 10^{-3}$ &number of iterations : $l \lesssim 4$  \\ \hline
 Auto-corr. time  && $\tau_{\rm int} \simeq 2$ for ${\rm Re} S[z] $\\ 
                          &&$\tau_{\rm int} \simeq 3$ for $\phi_z$ \\
\hline
\end{tabular}
\end{center}
\label{table:parameters-symmetric-phase}
\end{table}%

%

%\vspace{5em}
%\newpage
%{$\phantom{A}$}
%\newpage

\begin{figure}[htbp]
\begin{center}
%\vspace{3em}
\includegraphics[width=10.cm]{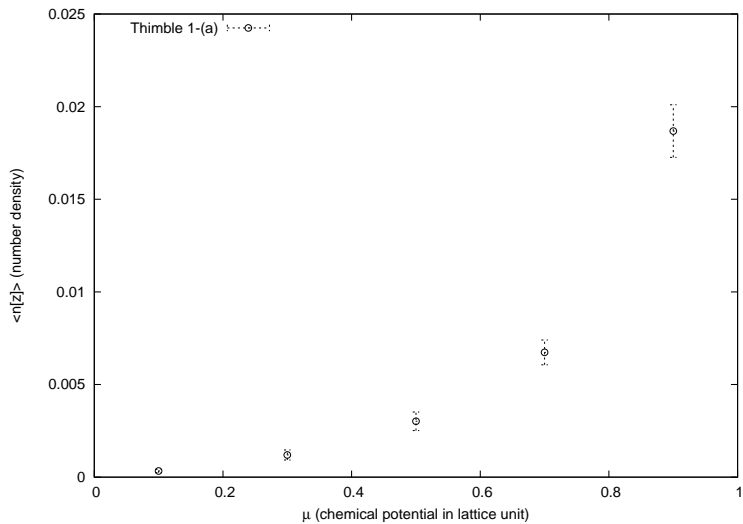}
\end{center}
\caption{The expectation values of $n[z]$ evaluated on the thimble 1-(a) ($\mu < \tilde \mu_c$). The errors are those estimated by the jack-knife method.}
\label{fig:number-deinsity-in-symmetric-phase}
\end{figure}

We have made measurements of $n[z]$ and ${\rm e}^{i \phi_z}$ using 300 trajectories 
out of 4,250 with separations of 10, 
discarding the first 1,250 for thermalization. 
The numerical results of $\langle {\rm e}^{i \phi_z} \rangle^\prime_{\mathcal{J}_{\text{vac}} }$, listed in  
table~\ref{table:residula-phase-symmetric-phase}, suggest  that the reweighting would work for all the given values of $\mu \, (< \tilde \mu_c)$.
The result of $\langle n[z] \rangle_{\mathcal{J}_{\text{vac}} }$, 
based on the formula eq.~(\ref{eq:observables-on-vacuum-thimble-residual-phase-quenched}), 
is shown in fig.~\ref{fig:number-deinsity-in-symmetric-phase}. 
The errors are those estimated by the jack-knife method.

\begin{table}[b]
\caption{Averages of the residual phase factors. The errors are statistical ones.}
\begin{center}
\begin{tabular}{|c|c|}
\hline
$\mu$ & 
%$\langle o[z] \rangle_{\mathcal{J}_\sigma}$
$\langle {\rm e}^{i \phi_z} \rangle^\prime_{\mathcal{J}_{\text{vac}} }$ 
%&
%$\langle n[z] \rangle_{\mathcal{J}_{\text{vac}}}$ &
%$\langle {\rm e}^{i \phi_z} \, n[z]  \rangle_{\mathcal{J}_{\text{vac}}}$ &
%$\langle n[z] \rangle$
\\ 
\hline
0.1 & (9.99e-01, -1.15e-03) $\pm$ (5.7e-02, 7.4e-04) 
\\ 
0.3 & (9.99e-01, -1.03e-03) $\pm$ (5.7e-02, 2.1e-03)
\\
0.5 & (9.98e-01, -2.68e-03) $\pm$ (5.7e-02, 3.3e-03)
\\
0.7 & (9.97e-01, \,5.24e-04) $\pm$ (5.7e-02, 4.3e-03)
\\
0.9 & (9.94e-01, -7.40e-03) $\pm$ (5.7e-02, 5.9e-03)\\
%0.1 & (9.999e-01, -1.156e-03) $\pm$ (5.7e-02, 7.4e-04) 
%\\ 
%0.3 & (9.992e-01, -1.035e-03) $\pm$ (5.7e-02, 2.1e-03)
%\\
%0.5 & (9.983e-01, -2.680e-03) $\pm$ (5.7e-02, 3.3e-03)
%\\
%0.7 & (9.970e-01, \,5.249e-04) $\pm$ (5.7e-02, 4.3e-03)
%\\
%0.9 & (9.946e-01, -7.405e-03) $\pm$ (5.7e-02, 5.9e-03)\\
\hline
\end{tabular}
\end{center}
\label{table:residula-phase-symmetric-phase}
\end{table}%

%\newpage
%{$\phantom{A}$}
\newpage

\subsection{Thimble 2-(b) for $\mu > \tilde \mu_c$}

On the other hand, 
when applied to the thimble 2-(b) for $\mu > \tilde \mu_c$, 
the algorithm in section~\ref{sec:HMC-on-Lefschetz-thimble} 
requires a few modifications in the parametrization of the thimble. 
This is because 
%the thimble of 2-(b) is of dimension 2$V$-1 
the thimble of 2-(b) has the critical region  of dimension one %2$V$-1 
and there appears a zero mode $\kappa^0 (=0)$ 
which corresponds to the degrees of freedom in the parameter $\theta$ 
(i.e. the zero-momentum modes of the Nambu-Goldstone boson $\pi$). 
In fact, the asymptotic solution to the flow equation in this case is given by
\begin{eqnarray}
\label{eq:asymptotic-solution-phi4}
z_a(x ; t) &\simeq& R_{ab}(\theta) \biggl\{ 
\delta_{b1}  \phi_0
+\sum_{\beta=1}^{2V-1} v_b(x)^\beta \, \exp( \kappa^\beta t) \,  e^\beta \biggr\} \qquad (t \ll 0), %\\
%%\, ; \quad
%%{\scriptstyle \sum_{\beta=1}^{2V-1} } e^\beta e^\beta = 2V-1, \nonumber\\
%V_a(x ; t)^\beta &=& R_{ab}(\theta) \,  v_b(x)^\beta \, \exp( \kappa^\beta t) \quad (\beta=1,\cdots, 2V-1), \\
%V_a(x ; t)^0 &=& R_{ab}(\theta) \,  v_b(x)^0, 
\end{eqnarray}
where the direction vector 
$e^\beta$ is  (2$V$-1)-dimensional and normalized as
$\sum_{\beta=1}^{2V-1} e^\beta e^\beta = \text{(2$V$-1)}$, 
%$v_b(x)^0 = \delta_{b2} / \sqrt{V}$, 
%$v_a(x)^1 = \delta_{a1} / \sqrt{V}$,\footnote{
%See appendix for the expressions of$\{ v_a(x)^\beta \}$ and  $\{\kappa^\beta\}$
%for $\beta=2,\cdots, 2V-1$.}
and 
$R(\theta) \in \text{O(2)}$: $R_{11}=R_{22}=\cos \theta$ and $R_{21}=-R_{12}=\sin \theta$.\footnote{See appendix for the expressions of $v_a(x)^\beta$ and  $\kappa^\beta$
for $\beta=0, 1,\cdots, 2V-1$.}
As for the variation $\delta z_a(x;t)$, it follows  that 
%And it follows for the variation $\delta z_a(x;t)$ that
\begin{equation}
\delta z_a(x ; t) = V_a(x ; t)^0 \big(\phi_0 \sqrt{V}  \delta \theta \big) +
\sum_{\beta=1}^{2V-1} V_b(x ; t)^\beta (\delta e^\beta + \kappa^\beta e^\beta \delta t ).
\end{equation}
We regard $\theta$ as a dynamical variable in the molecular dynamics.
According to 
the equations of motion eqs.~(\ref{eq:equations-of-motion-with-Lagrange-multipliers-z}) 
and (\ref{eq:equations-of-motion-with-Lagrange-multipliers-w}), it obeys
$\phi_0 \sqrt{V}  \dot{\theta} = (w)^0$ and $(\dot{w})^0 = 0$ because $\kappa^0=0$.
%When $\mu \gtrsim \tilde \mu_c$, however,  $\kappa^1 = 2 \lambda_0 \langle \phi \rangle^2$ can be very small
%and, due to critical fluctuations,  the component $e^1$ can dominate the normalized vector 
%$e^\beta$, implying that the factor ``$e^1 \exp(\kappa^1 t)$'' is not a small number unless
%$t$ assumes to be a very large negative value. And this can invalidate the linear approximation to the flow equations.
%

%Moreover, 
Furthermore, 
when $\mu$ is close to $\tilde \mu_c$ ($\mu \gtrsim \tilde \mu_c$), 
%there appears
%moreover, 
the lowest lying non-zero mode 
with $\kappa^1= 2 \lambda_0 \phi_0^2$ and $v_a(x)^1 = \delta_{a1} / \sqrt{V}$
%, etc,, in the non-zero modes $\{ \kappa^\beta \}$ $(\beta=1,\cdots, 2V-1)$ 
(i.e. the zero-momentum mode of the scalar boson $\sigma$) tends to be very light\footnote{Here we assume the lattice size $L$ is relatively small. For a large $L$, there also appear light non-zero momentum modes of the scalar and Nambu-Goldstone bosons.}
and, due to critical fluctuations,\footnote{The critical point of the second-order phase transition in this system is $\mu_c \simeq 1.15$ ($\gtrsim \tilde \mu_c$) for $\kappa=1, \lambda=1$, as shown  in \cite{Aarts:2008wh,Aarts:2009hn}.}
the component $e^1$ can dominate the direction vector $e^\beta$. 
%$(\beta=1,\cdots, 2V-1)$ normalized as $\sum_\beta^{2V-1} e^\beta e^\beta = 2V-1$.
%
This implies that the factor $\exp(\kappa^1 t) e^1$ 
in the asymptotic solution eq.~(\ref{eq:asymptotic-solution-phi4}) 
%eq.~(\ref{eq:flow-in-asymptotic-region}) (or the initial condition eq.~(\ref{eq:initial-condition-z}))
is not a small number unless $t$ (or $t_0$) assumes %to be 
a very large negative value, 
and this can invalidate the linear approximation to the flow equations.\footnote{One should  also note 
the fact that the truncation errors in the linear approximation are of order $\lambda_0 z^3$ for the critical points 
1-(a) ($\mu < \tilde \mu_c$), but of order $\lambda_0 \phi_0 (z-\phi_0)^2$ 
for the critical point 2-(b) ($\mu > \tilde \mu_c$).   
For the latter case, it is relatively hard to reach the asymptotic region.}
To improve this situation, we note that for the {\em global} flow mode $z_a(x ; t)=z_a(t)$, 
the flow equation reads
 \begin{eqnarray}
&& \frac{d}{dt} z_a(t) \, \,  \, \, 
=  \left. \bar{\partial}_{ax} \bar S[ \bar z ] \right\vert_{z_a(x ; t)=z_a(t)} \nonumber\\
&& \qquad \qquad \, = 
  \lambda_0 \big(\bar{z}_b(t) \bar{z}_b(t)-\phi_0^2 \big) \bar{z}_a(t), % \\
%                            \qquad\qquad (a=1,2)
%&&\nonumber\\
%&& \frac{d}{dt} V_{a}(t)^\beta 
%= \left. \bar{\partial}_{ax} \bar{\partial}_{by} \bar S[ \bar z ] \right\vert_{z_a(x ; t)=z_a(t)} \, \bar{V}_{b}(t)^\beta 
%\nonumber\\
%&& \qquad \qquad \, = 
%\lambda_0 \big(\bar{z}_b(t) \bar{z}_b(t)-\langle \phi \rangle^2 \big) \bar{V}_a(t)^\beta
%+ 2\lambda_0 \bar{z}_a(t) \,  \bar{z}_b(t) \bar{V}_b(t)^\beta , 
%%\quad (\beta=0,1), 
\end{eqnarray}
and the exact solution to the {\em non-linear} flow equation is obtained explicitly as  
\begin{eqnarray}
z_a(t) &=&   
R_{ab}(\theta) %\biggl\{ 
\delta_{b1} \, 
\frac{ \phi_0  }
{ \sqrt{ 1- \frac{2}{ \sqrt{V} \phi_0 } \, e^1 
                                                                          \, \exp(\kappa^1 t ) }}.
% , \\
%{\Big( 1- \frac{2}{ \langle \phi \rangle \sqrt{V} } \, e^1 \, \exp(\kappa^1 t ) \Big)^{1/2}}
%+\sum_{\beta=2}^{2V-1} v_b(x)^\beta \, \exp( \kappa^\beta t) \,  e^\beta 
%\biggr\}
%, \\
%V_a(t)^0 &=& R_{ab}(\theta) \,  v_b^0 \, 
%\frac{1}
%{ \sqrt{ 1- \frac{2}{ \langle \phi \rangle \sqrt{V} } \, e^1 \, \exp(\kappa^1 t ) }} ,  \\
%%{\Big( 1- \frac{2}{ \langle \phi \rangle \sqrt{V} } \, e^1 \, \exp(\kappa^1 t ) \Big)^{1/2}} , \\
%V_a(t)^1 &=& R_{ab}(\theta) \,  v_b^1 \, 
%\frac{\exp(\kappa^1 t ) }
% {\Big( 1- \frac{2}{ \langle \phi \rangle \sqrt{V} } \, e^1 \, \exp(\kappa^1 t ) \Big)^{3/2}} .
\end{eqnarray}
Here the allowed range of $t$ is $[-\infty, t^\ast]$ where $t^\ast =\ln(\sqrt{V} \phi_0 / 2 e^1)/\kappa^1$, 
%Here $e^1$, as an integration constant, 
%determines the allowed range of $t$ as $[-\infty, t^\ast]$ where 
%$t^\ast =\ln(\sqrt{V} \phi_0 / 2 e^1)/\kappa^1$, 
and $e^1$ takes a value in the range
$[-\infty,  {e^1}^\ast ]$ where ${e^1}^\ast= \sqrt{V} \phi_0 \, \exp(-\kappa^1 t_0 ) /2$ for $t=t_0 (\ll 0)$ fixed.
%Here $c$, as an integration constant, takes a value in the range 
%$\big[-\infty,  \langle \phi \rangle \sqrt{V} \, \exp(-\kappa^1 t ) /2 \big]$ for a fixed $t$.
%This suggests that we adopt 
This leads us to adopt the following asymptotic form for $t \ll 0$, 
\begin{equation}
z_a(x ; t) \simeq R_{ab}(\theta) \left\{ 
\delta_{b1} 
\frac{ \phi_0  }{ \sqrt{ 1- \frac{2}{ \sqrt{V} \phi_0 } \, e^1 \, 
\exp(\kappa^1 t ) }}
+\sum_{\beta=2}^{2V-1} v_b(x)^\beta \, \exp( \kappa^\beta t) \,  e^\beta \right\} , 
\end{equation}
where the direction vector $e^\beta$ is normalized as $\sum_{\beta=2}^{2V-1} e^\beta e^\beta = \text{2$V$-2}$ 
excluding $e^1$. 
Accordingly, for the tangent vectors, we adopt the following asymptotic forms  for $t \ll 0$,
\begin{eqnarray}
\label{eq:asymptotic-form-V-0}
V_a(x; t)^0 &\simeq& R_{ab}(\theta) \,  v_b(x)^0 \, 
\frac{1}
{ \sqrt{ 1- \frac{2}{ \sqrt{V} \phi_0 } \, e^1 \, \exp(\kappa^1 t ) }} , \\
\label{eq:asymptotic-form-V-1}
V_a(x; t)^1 &\simeq& R_{ab}(\theta) \,  v_b(x)^1 \, 
\frac{\exp(\kappa^1 t ) }
 {\Big( 1- \frac{2}{ \sqrt{V} \phi_0 } \, e^1 \, \exp(\kappa^1 t ) \Big)^{3/2}} , \\
\label{eq:asymptotic-form-V-beta}
V_a(x ; t)^\beta &\simeq& R_{ab}(\theta) \,  v_b(x)^\beta \, \exp( \kappa^\beta t) \qquad\qquad (\beta=2,\cdots, 2V-1), 
\end{eqnarray}
where $v_a(x)^0 = \delta_{a2} / \sqrt{V}$.\footnote{
The tangent vectors $V_a(x; t)^0$ and $V_a(x; t)^1$ in 
(\ref{eq:asymptotic-form-V-0}) and (\ref{eq:asymptotic-form-V-1}), respectively 
are indeed the exact solutions to the flow equations with the global flow mode $z_a(x ; t)=z_a(t)$:
 \begin{eqnarray}
%&& \frac{d}{dt} z_a(t) \, \,  \, \, 
%=  \left. \bar{\partial}_{ax} \bar S[ \bar z ] \right\vert_{z_a(x ; t)=z_a(t)} \nonumber\\
%&& \qquad \qquad \, = 
%  \lambda_0 \big(\bar{z}_b(t) \bar{z}_b(t)-\langle \phi \rangle^2 \big) \bar{z}_a(t), % \\
%%                            \qquad\qquad (a=1,2)
%%&&\nonumber\\
\frac{d}{dt} V_{a}(x; t)^\beta 
&=& \left. \bar{\partial}_{ax} \bar{\partial}_{by} \bar S[ \bar z ] \right\vert_{z_a(x ; t)=z_a(t)} \, \bar{V}_{b}(y; t)^\beta 
\nonumber\\
&=& 
K_0 \Delta_{ab}
%\big[ 
% \{ \nabla_k \nabla_k^\ast + \cosh(\mu) \nabla_0 \nabla_0^\ast \} \delta_{ab}
%  -i \sinh(\mu) (\nabla_0 + \nabla_0^\ast)  \epsilon_{ab} 
%\big] 
\bar{V}_b(x; t)^\beta 
%\nonumber\\
%&& 
 + \lambda_0 \big(\bar{z}_b(t) \bar{z}_b(t)- \phi_0^2 \big) \bar{V}_a(x; t)^\beta
 + 2\lambda_0 \bar{z}_a(t) \,  \bar{z}_b(t) \bar{V}_b(x; t)^\beta, \nonumber
%&=& \big[ 
%K_0 \{ \nabla_k \nabla_k^\ast + \cosh(\mu) \nabla_0 \nabla_0^\ast \} 
%+\lambda_0 \big(\bar{z}_b(t) \bar{z}_b(t)- \phi_0^2 \big) \big] \bar{V}_a(x; t)^\beta 
%\nonumber \\
%&& + \big[ K_0 \{ -i \sinh(\mu) (\nabla_0 + \nabla_0^\ast) \} \epsilon_{ab} 
%+ 2\lambda_0 \bar{z}_a(t) \,  \bar{z}_b(t) \big]\bar{V}_b(x; t)^\beta. \nonumber
%%\quad (\beta=0,1), 
\end{eqnarray}
where $\Delta_{ab}=
 \{ \nabla_k \nabla_k^\ast + \cosh(\mu) \nabla_0 \nabla_0^\ast \} \delta_{ab}
  -i \sinh(\mu) (\nabla_0 + \nabla_0^\ast)  \epsilon_{ab}$.
The similar exact solutions for $V_a(x; t)^\beta$ ($\beta=2,\cdots,2V-1$) can be worked out, but the results turns out to be involved. We therefore adopt the simpler solutions to the linearized flow equation as 
in (\ref{eq:asymptotic-form-V-beta}), although the consistency in the linear approximation is lost.
}

Using the algorithm with the above modifications, we have generated 
$11,250$ trajectories for each value $\mu=1.0, 1.1, 1.2, 1.3$, and $1.5$
with the parameters 
listed in table~\ref{table:parameters-broken-phase}.
In this case, each trajectory has the length $\tau_{\rm traj}=0.3$ and obtained in the number of steps
$n_{\rm step}=30 \, (\mu=1.0, 1.1)$ and $10 \, (\mu=1.2, 1.3, 1.5)$. 
In solving the flow equations, the parameters are chosen as $t_0 =-3.0$ and $n_{\rm lefs}=100$.
In the course of the updates, we have found that
$t^\prime \in [ 2.5, 3.5]$ and $h=t^\prime/n_{\rm lefs} \simeq 0.03$ most of the time, 
and the solutions satisfy the bounds,  
$ \vert {\rm Im} ( S[z]-S[z_{\rm vac}])  \vert \lesssim 5.0 \times 10^{-2}$ and 
$\| \bar \partial \bar S - V^\alpha \kappa^\alpha e^\alpha \|^2 /2V \lesssim 3.0 \times 10^{-2}$. 
In solving the constraint in the molecular dynamics, 
the fixed-point method converges with iteration numbers 
$l \le 6 \, (\mu=1.0), 14 \, (\mu=1.1), 4 \, (\mu=1.2, 1.3, 1.5)$ 
for the step sizes 
$\Delta \tau =\tau_{\rm traj}/ n_{\rm step}=0.01 \, (\mu=1.0, 1.1), 0.03 \, (\mu=1.2, 1.3, 1.5)$ 
and the bound $\epsilon^\prime = \sqrt{10} \times 10^{-3}$.
It has occurred twice for $\mu=1.0$ and once for $\mu=1.1$ that the fixed point method failed to converge. For such trajectories, the momenta have been re-refreshed and the molecular dynamics has been re-started.\footnote{
As far as we understand, these failures have occurred due to our implementation of the algorithm.
The asymptotic solution is in the form of the ``polar decomposition'' as $z_a \simeq R_{a 1}(\theta) \rho$, where 
$\rho = \phi_0 / \sqrt{ 1- 2 e^1 {\rm e}^{\kappa^1 t} / \phi_0 \sqrt{V} }$. 
The factor $\rho$ can be rather small for $\mu \gtrsim \tilde \mu_c$, and 
it can even be negative in the updates with a finite step size. 
In such a case, 
one needs to do a coordinate transformation such as 
$( \rho, \theta ) \rightarrow ( - \rho, \theta+\pi)$.
This procedure is in fact neglected in our implementation, and we have  instead managed with the reduced step size $\Delta \tau = 0.01$ $(\mu=1.0, 1.1)$.
}
%The acceptance rates are $\simeq$ 0.99 in averages.     
%The integrated auto-correlation times are estimated 
%as $\tau_{\rm int} \simeq 14 \, (\mu=1.0, 1.1), 10\, (\mu=1.2, 1.3, 1.5)$ for ${\rm Re} S[z] $
%and  $\tau_{\rm int} \simeq 14 \, (\mu=1.0), 28 \, (\mu=1.1), 15\, (\mu=1.2, 1.3, 1.5) $ for $\phi_z$.

%
\begin{table}[htbp]
\caption{Simulation parameters for the thimble 2-(b) ($\mu > \tilde \mu_c$)}
\begin{center}
\begin{tabular}{| l | l | l |}
\hline
& Parameters & Resulting conditions \\ \hline
Thimble & $t_0 =-3.0$ &  $ \vert {\rm Re} \big( S[z(t_0)] - S[z_{\rm vac}] \big)\vert \lesssim 2.0\times 10^1 $ \\ 
%(Solving flow eqs.) 
& $n_{\rm lefs}=100$ & $ \vert {\rm Im} (S[z]-S[z_{\rm vac}]) \vert \lesssim 5.0 \times 10^{-2}$ \\
&$h=t^\prime/n_{\rm lefs} \simeq 0.03$& 
$\| \bar \partial \bar S - V^\alpha \kappa^\alpha e^\alpha \|^2 /2V \le 3.0 \times 10^{-2}$ \\ \hline
MD & $\tau_{\rm traj}=0.3$& $t^\prime \in [ 2.5, 3.5]$ \\ 
%(Solving constraint)  
&$n_{\rm step}=10$, $30$ ($\mu=1.0, 1.1$) & $\Delta H \lesssim 0.05 $\\
   &$\Delta \tau=0.03$, $0.01$ ($\mu=1.0, 1.1$)& Acceptance rate $\simeq 0.99$\\
   &$\epsilon^\prime = \sqrt{10} \times 10^{-3}$ 
                                                &$l \lesssim 4$,  $6$ ($\mu=1.0$),  $14$ ($\mu=1.1$)  \\ \hline
 Auto-corr. time  &(for ${\rm Re} S[z] $)& $\tau_{\rm int} \simeq 10$, $14$ ($\mu=1.0, 1.1$)\\ 
                          &(for $\phi_z$)&$\tau_{\rm int} \simeq 15$, $14$ ($\mu=1.0$), $28$ ($\mu=1.1$) \\
\hline
\end{tabular}
\end{center}
\label{table:parameters-broken-phase}
\end{table}%

\begin{figure}[htbp]
\begin{center}
\vspace{2em}
\includegraphics[width=10.cm]{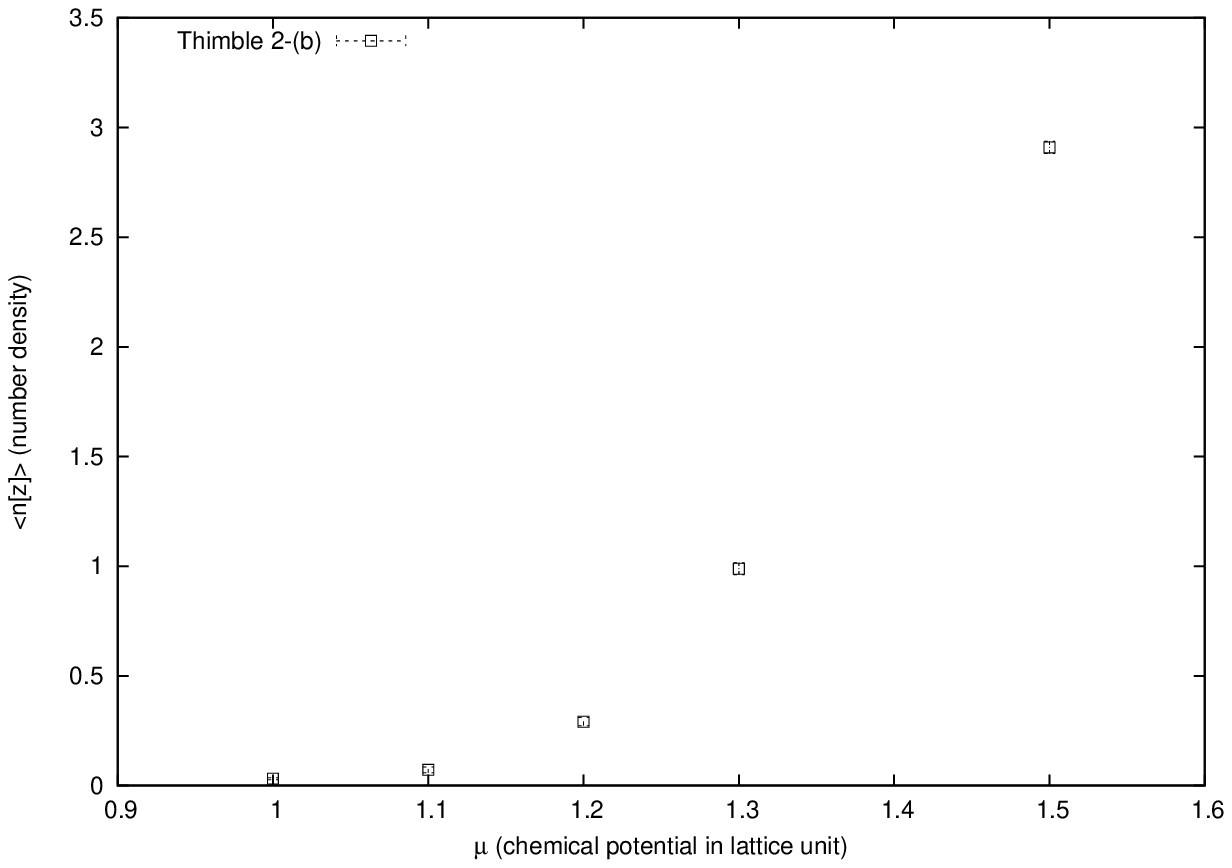}
\end{center}
\caption{The expectation values of $n[z]$ evaluated on the thimble 2-(b) ($\mu > \tilde \mu_c$). The errors are those estimated by the jack-knife method.}
\label{fig:number-deinsity-in-broken-phase}
\end{figure}

%\newpage
We have made measurements of $n[z]$ and ${\rm e}^{i \phi_z}$ using 1,000 trajectories out of 11,250 
with separations of 10, 
discarding the first 1,250 for thermalization. 
The numerical result of $\langle {\rm e}^{i \phi_z} \rangle^\prime_{\mathcal{J}_{\text{vac}} }$, listed in  
table~\ref{table:residula-phase-broken-phase}, suggests again 
that the reweighting would work for all the given values 
of $\mu \, (> \tilde \mu_c)$.
The result of $\langle n[z] \rangle_{\mathcal{J}_{\text{vac}} }$, 
based on the formula eq.~(\ref{eq:observables-on-vacuum-thimble-residual-phase-quenched}), 
is shown in fig.~\ref{fig:number-deinsity-in-broken-phase}. 
The errors are those estimated by the jack-knife method.

\begin{table}[htdp]
\caption{Averages of the residual phase factor. The errors are statistical ones.}
\begin{center}
\begin{tabular}{|c|c|}
\hline
$\mu$ & 
$\langle {\rm e}^{i \phi_z} \rangle^\prime_{\mathcal{J}_{\text{vac}} }$ 
\\ 
\hline
1.0 & (9.94e-01, -8.77e-03) $\pm$ (3.1e-02, 3.1e-03) 
\\ 
1.1 & (9.94e-01, -3.21e-03) $\pm$ (3.1e-02, 3.4e-03)
\\
1.2 & (9.95e-01, -8.25e-04) $\pm$ (3.1e-02, 3.0e-03)
\\
1.3 & (9.97e-01, -3.08e-03) $\pm$ (3.1e-02, 2.2e-03)
\\
1.5 & (9.99e-01, -1.06e-03) $\pm$ (3.1e-02, 1.0e-03)\\
\hline
\end{tabular}
\end{center}
\label{table:residula-phase-broken-phase}
\end{table}%

\newpage
\subsection{A comparison to the results of the complex Langevin simulations}

\begin{figure}[dht]
\begin{center}
\includegraphics[width=11cm]{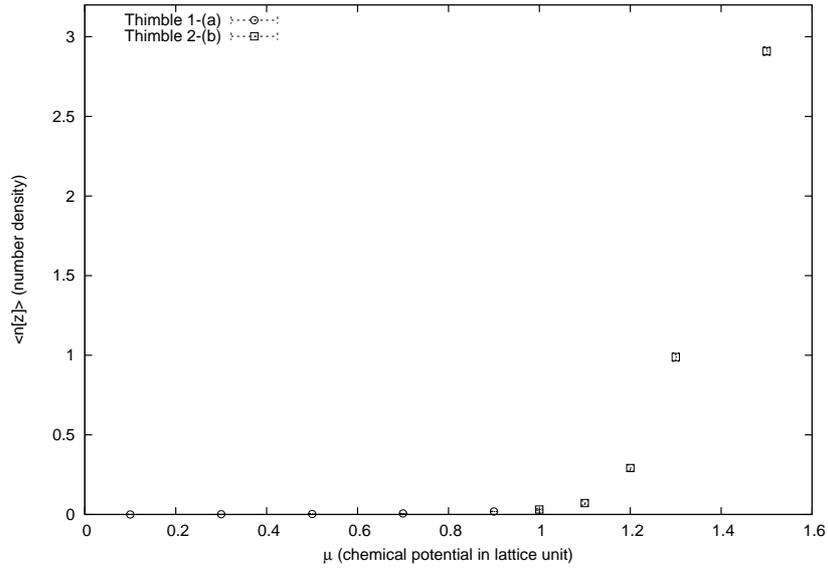}
\end{center}
\caption{The expectation values of $n[z]$ evaluated on both thimbles, 
1-(a) for $\mu < \tilde \mu_c$ and 2-(b) for $\mu > \tilde \mu_c$. The errors are those estimated by the jack-knife method.}
\label{fig:number-deinsity-in-both-phases}
\end{figure}

In fig.~\ref{fig:number-deinsity-in-both-phases}, the results of 
$\langle n[z] \rangle_{\mathcal{J}_{\text{vac}} }$ on the two thimbles, 1-(a) for $\mu < \tilde \mu_c$ 
and 2-(b) for $\mu > \tilde \mu_c$, are shown together.  
The numerical data are summerized in table~\ref{table:data-of-number-density}.

\begin{table}[htbp]
\caption{Numerical data of the expectation values of $n[z]$}
\begin{center}
\begin{tabular}{|c|c|c|c|}
\hline
$\mu$ & 
{\rm Re} $\langle n[z] \rangle_{\mathcal{J}_{\text{vac}}}$ (j.-k. error)&
{\rm Re} $\langle {\rm e}^{i \phi_z} \, n[z]  \rangle^\prime_{\mathcal{J}_{\text{vac}}}$ &
{\rm Re} $\langle n[z] \rangle^\prime_{\mathcal{J}_{\text{vac}}}$ 
\\ 
\hline
0.1 & 3.34e-04 \, (9.2e-05) & 3.35e-04 & 2.15e-04 \\
0.3 & 1.20e-03 \, (2.7e-04) & 1.19e-03 & 8.56e-04 \\
0.5 & 3.02e-03 \, (5.0e-04) & 3.01e-03 & 2.44e-03 \\
0.7 & 6.74e-03 \, (6.7e-04) & 6.71e-03 & 5.91e-03 \\
0.9 & 1.89e-02 \, (1.4e-03) & 1.85e-02 & 1.73e-02 \\
1.0 & 3.14e-02 \, (4.3e-03) & 3.12e-02 & 3.00e-02 \\
1.1 & 7.17e-02 \, (1.3e-02) & 7.12e-02 & 7.01e-02 \\
1.2 & 2.92e-01 \, (1.8e-02) & 2.90e-01 & 2.90e-01 \\
1.3 & 9.88e-01 \, (2.6e-02) & 9.85e-01 & 9.87e-01 \\
1.5 & 2.91e-00 \, (2.7e-02) & 2.90e-00 & 2.90e-00 \\
\hline
\end{tabular}
\end{center}
\label{table:data-of-number-density}
\end{table}%

%\begin{table}[htdp]
%\caption{Numerical data of the expectation values of $n[z]$}
%\begin{center}
%\begin{tabular}{|c|c|c|c|}
%\hline
%$\mu$ & 
%{\rm Re} $\langle n[z] \rangle$ (j.-k. error)&
%{\rm Re} $\langle {\rm e}^{i \phi_z} \, n[z]  \rangle_{\mathcal{J}_{\text{vac}}}$ &
%{\rm Re} $\langle n[z] \rangle_{\mathcal{J}_{\text{vac}}}$ 
%\\ 
%\hline
%0.1 & 3.34e-04 \, (9.2e-05) & 3.35e-04 & 2.15e-04 \\
%0.3 & 1.20e-03 \, (2.7e-04) & 1.19e-03 & 8.56e-04 \\
%0.5 & 3.01e-03 \, (5.0e-04) & 3.01e-03 & 2.44e-03 \\
%0.7 & 6.73e-03 \, (6.7e-04) & 6.71e-03 & 5.91e-03 \\
%0.9 & 1.86e-02 \, (1.4e-03) & 1.85e-02 & 1.73e-02 \\
%1.0 & 3.05e-02 \, (3.1e-03) & 3.04e-02 & 2.02e-02 \\
%1.1 & 8.34e-02 \, (1.2e-02) & 8.29e-02 & 8.19e-02 \\
%1.2 & 2.98e-01 \, (2.9e-02) & 2.97e-01 & 2.97e-01 \\
%1.3 & 9.74e-01 \, (2.4e-02) & 9.72e-01 & 9.73e-01 \\
%1.5 & 2.91e-00 \, (3.0e-02) & 2.91e-00 & 2.91e-00 \\
%\hline
%\end{tabular}
%\end{center}
%\label{table:data15-of-number-density}
%\end{table}%

\begin{figure}[dht]
\begin{center}
\includegraphics[width=11.5cm]{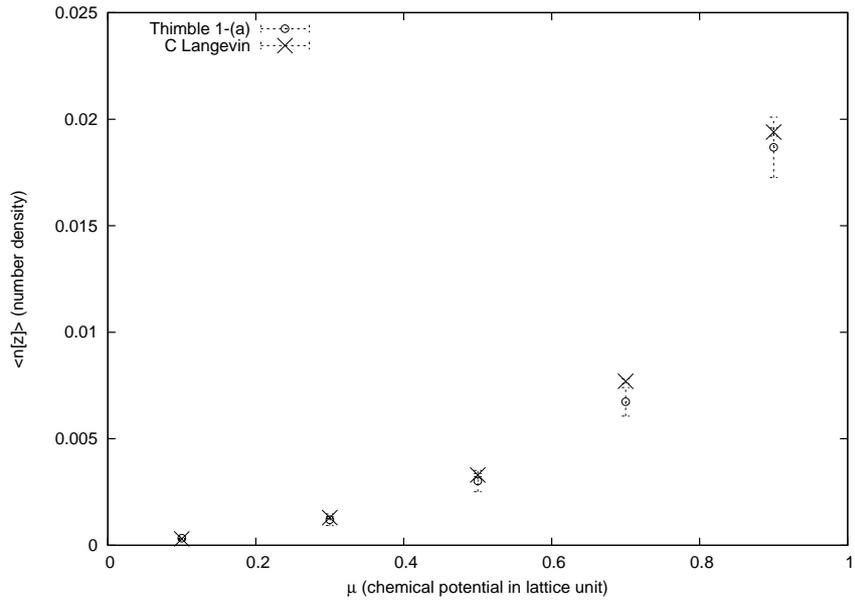}

\vspace{4em}
\includegraphics[width=11.5cm]{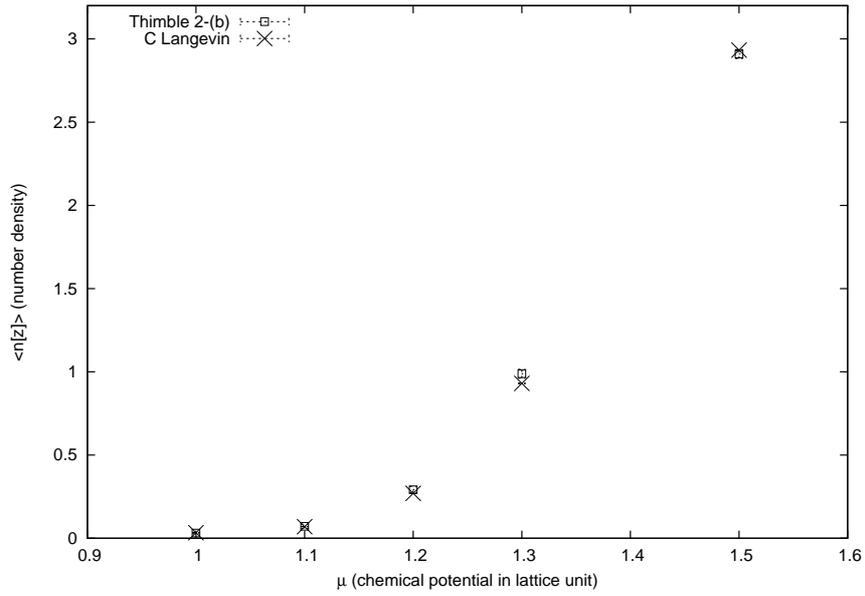}
\end{center}
\caption{
The expectation values of $n[z]$ evaluated 
by the complex Langevin simulations in comparison with those by the hybrid Monte Carlo.}
\label{fig:number-deinsity-in-comparison-with-CLE}
\end{figure}

It is instructive to compare our numerical results with those obtained by the complex Langevin equation\cite{Aarts:2008wh} and the dual variable method\cite{Gattringer:2012df,Mercado:2012ue,Gattringer:2012jt}.
We have reproduced the expectation values of $n[z]$ 
through the complex Langevin simulations with the step size $\epsilon =5.0 \times 10^{-5}$, 
samping 10,000 configurations with separation of 500 out of $5.0 \times 10^{6}$ timesteps. 
These results are shown 
in fig.~\ref{fig:number-deinsity-in-comparison-with-CLE} with our results by the hybrid Monte Carlo. 
The two sets of the results are in agreement within the statistical errors, 
except for $\mu=0.7, 1.2, 1.3$,  and overall, they are consistent with each other.
%This implies that our hybrid Monte Carlo method is working at least for the thimbles, 1-(a) for 
%$\mu < \tilde \mu_c$ and 2-(b) for $\mu > \tilde \mu_c$, in the $L=4$ lattice, and also the approximation 
%based on eq.~(\ref{eq:observables-on-vacuum-thimble}) is rather good for 
%all the given values of $\mu$.

\newpage
{$\phantom{A}$}
\newpage
{$\phantom{A}$}
\newpage

\section{Summary and Discussion}
\label{sec:summary-discussion}

In this article, we have introduced the hybrid Monte Carlo algorithm which is applicable to   
lattice models defined on Lefschetz thimbles. We  have tested the algorithm 
in the $\lambda \phi^4$  model with the couplings $\kappa=1.0$ and $\lambda=1.0$, the chemical potential $\mu \in [0.0, 1.5]$($\tilde \mu_c \simeq 0.962$), and the lattice size $L=4$. 
We have found that 
the algorithm can indeed be applied to the thimbles associated with the classical vacua 
%1-(a) for $\mu < \tilde \mu_c$ and 2-(b) for $\mu > \tilde \mu_c$, 
and can produce the expectation values of the number density, $n[z]$,  
which are consistent with those obtained by the other 
methods\cite{Aarts:2008wh,Gattringer:2012df,Mercado:2012ue,Gattringer:2012jt}. 
In particular,  we have shown that 
the residual sign factors, ${\rm e}^{i \phi_z} = \det V_z / \vert \det V_z  \vert$,  
average to not less than $0.99$,  and can be safely included by reweighting
for all the values of $\mu$ studied within the range $[0.0, 1.5]$.
This result is in sharp contrast to 
the fact that 
the phase-quenched Monte Carlo method based on the real part of the action, ${\rm Re} S[x]$,  fails
because the averages of the exponent of the imaginary part of the action, 
$\langle {\rm e}^{- i {\rm Im} S[x]} \rangle$, get vanishingly small  for $\mu \gtrsim \tilde \mu_c$ even at the lattice size $L=4$\cite{Aarts:2008wh,Cristoforetti:2013wha}.

As a next step, we certainly need  
to examine in detail the systematic errors in the hybrid Monte Carlo method, 
in particular, those  in defining the asymptotic regions of the thimbles and 
in neglecting the possible contributions of the other thimbles. In this respect, it is 
somewhat surprising to observe the agreement of the two set of the results shown in fig.~\ref{fig:number-deinsity-in-comparison-with-CLE}, in particular, for the values of $\mu$ close to $\tilde \mu_c$,  because
we expect that the contributions 
of the other thimbles such as that associated with 2-(a) would become important there.
We should also extend the study of the residual sign problem in the 
$\lambda \phi^4$  model to larger lattice sizes.
The numerical cost per trajectory in the algorithm scales as 
${\cal O}(V^2 n_{\rm lefs} \times n_{\rm step})$ in the computation of the tangent vectors 
$\{ V_z^\alpha \}$ $(\alpha =1,\cdots, 2V)$ and 
as ${\cal O}(V^3 \times n_{\rm step})$ in the computation 
of the inverse ${V_z}^{-1}$ and the determinant $\det V_z$, and 
it would be challenging for the large lattice sizes.\footnote{
For this, 
%since the computation of the tangent vectors 
%$\{ V_z^\alpha \}$ $(\alpha =1,\cdots, n)$ get numerically   demanding further, 
it would be essential to execute the computation of the tangent vectors 
%$\{ V_z^\alpha \}$ $(\alpha =1,\cdots, n)$
in parallel using GPUs (cluster).
} 
A study on these points will be reported in a forth coming paper.

For a future study, it would be interesting to apply the hybrid Monte Carlo method to lattice QCD at finite density. Results in this approach, even at small lattices,   
would serve as a cross check of the results
obtained recently in the complex Langevin approach\cite{Sexty:2013ica}.

\begin{acknowledgments}
The authors would like to thank Koji~Hukushima for enlightening conversations.
H.F., Y.K., and T.S. are grateful to J.~Bloch and F.~Bruckmann for their hospitality at the International Workshop on the Sign Problem in QCD and Beyond, and they are also grateful to L.~Scorzato for valuable discussions. The authors would also like to thank G.~Aarts for his comments and for sending us the numerical data of his results.
This work is supported in part by JSPS KAKENHI Grant Numbers 24540255 (H.F.), 25287049 (M.K.), 24540253 (Y.K.). 
D.H. and S.K. are supported in part by JSPS Research Fellowship
for Young Scientists.
\end{acknowledgments}
  
\newpage
\appendix
\section{Tangent vectors at the critical point of the thimble 2-(b)}

In this appendix, we give the explicit formulae of the tangent vectors $\{ v_a(x)^\alpha \}$ 
at the critical point of the thimble 2-(b). Let us label the tangent vectors by the set of indices
$\alpha = (k, \delta)$, where 
$k=(k_0, k_1, k_2, k_3)$ ($k_0=0,\cdots, L/2$; $k_i=0, \cdots, L-1$$(i=1,2,3)$) 
and $\delta=1,2,3,4$ ($\delta =1,3$ for $k=(0,0,0,0)$ and $(L/2,0,0,0)$).
Then $ \kappa(k,\delta)$ and $v_a(x; k,\delta)$ are given as follows:
\begin{enumerate}
\item $\kappa(k,1) = \Delta_\sigma  c^2 - \Delta_\pi s^2 + 2 S c s$: 
\begin{eqnarray}
v_a(x; k,1) &=&
(+1)
\sqrt{\frac{1}{L}}
 \left( \begin{array}{c} c \,  \cos(\frac{2\pi}{L} k_0 x_0) \\ -i s \, \sin(\frac{2\pi}{L} k_0 x_0) \end{array} \right) 
\, \prod_{i=1}^3 T_i(x_i; k_i) 
\end{eqnarray}

\item $\kappa(k,2) = \Delta_\sigma  c^2 - \Delta_\pi s^2 + 2 S c s$: 
\begin{equation}
v_a(x; k,2)
 =
(-i) 
\sqrt{\frac{1}{L}} \left( \begin{array}{c} i c \, \sin(\frac{2\pi}{L} k_0 x_0)  \\ - s \, \cos(\frac{2\pi}{L} k_0 x_0) \end{array} \right) 
\, \prod_{i=1}^3 T_i(x_i; k_i) 
\end{equation}

 \item $\kappa(k,3) = - \Delta_\sigma  s^2 + \Delta_\pi  c^2 + 2 S c s$ :
\begin{equation}
v_a(x; k,3)
 =
(+1)
\sqrt{\frac{1}{L}} \left( \begin{array}{c} i s \sin(\frac{2\pi}{L} k_0 x_0)  \\ c \cos(\frac{2\pi}{L} k_0 x_0) \end{array} \right) 
\, \prod_{i=1}^3 T_i(x_i; k_i) 
\end{equation}

 \item $\kappa(k,4) = - \Delta_\sigma  s^2 + \Delta_\pi  c^2 + 2 S c s$ :

\begin{equation}
v_a(x; k,4)
 =
(-i) 
\sqrt{\frac{1}{L}} \left( \begin{array}{c} s \cos(\frac{2\pi}{L} k_0 x_0)   \\ i c  \sin(\frac{2\pi}{L} k_0 x_0) \end{array} \right) 
\, \prod_{i=1}^3 T_i(x_i; k_i) 
\end{equation}

\end{enumerate}
where 
\begin{eqnarray}
T_i (x_i ; k_i) \equiv
\left\{
\begin{array}{ll} 
\sqrt{\frac{1}{L}}  &  (k_i=0)  \\
\sqrt{\frac{2}{L}}\cos(\frac{2\pi}{L} k_i x_i) & (k_i=1,\cdots, L/2-1) \\
\sqrt{\frac{1}{L}} (-1)^{x_i}  & (k_i = L/2) \\
\sqrt{\frac{2}{L}}\sin(\frac{2\pi}{L} k_i x_i) & (k_i=L/2+1,\cdots, L-1)
\end{array}
\right. 
\end{eqnarray}
($i=1,2,3)$, 

\begin{eqnarray}
\Delta_\pi(k) &=&(2 K_0)\left[  \sum_{i=1}^3 \Big(1- \cos(\frac{2\pi}{L} k_i ) \Big)
                          +\Big(1- \cos(\frac{2\pi}{L} k_0) \Big)\cosh(\mu) \right] , \\
\Delta_\sigma(k) &=& \Delta_\pi(k) + 2 \lambda_0 \langle \phi \rangle^2 ,  \\
S(k) &=& 2K_0 \, \sin(\frac{2\pi}{L} k_0) \sinh(\mu) , 
\end{eqnarray}
and
\begin{eqnarray}
c &=& \left[ \frac{\Delta_\sigma + \Delta_\pi}{2} + \sqrt{ \Big( \frac{\Delta_\sigma + \Delta_\pi}{2} \Big)^2+ S^2} \right]/N , \\
s &=& S / N , \\
N^2 
&=& 
\left[ \frac{\Delta_\sigma + \Delta_\pi}{2} + \sqrt{ \Big( \frac{\Delta_\sigma + \Delta_\pi}{2} \Big)^2+ S^2} \right]^2 + S^2 . 
\end{eqnarray}

\end{document}